%% file: template_isit24.tex
\documentclass[conference,letterpaper]{IEEEtran}

\input{macros}

\begin{document}
\title{Performance Analysis of Generalized Product Codes with Irregular Degree Distribution} 

\author{%
  \IEEEauthorblockN{Sisi Miao\textsuperscript{\dag}, Jonathan Mandelbaum\textsuperscript{\dag}, Lukas Rapp\textsuperscript{\ddag}, Holger Jäkel\textsuperscript{\dag}, and Laurent Schmalen\textsuperscript{\dag}}
  \IEEEauthorblockA{\textsuperscript{\dag}Karlsruhe Institute of Technology (KIT),
Communications Engineering Lab (CEL),
76187 Karlsruhe, Germany\\
\textsuperscript{\ddag}\emph{now with} Research Laboratory of Electronics (RLE), Massachusetts Institute of Technology, Cambridge, MA 02139 USA\\
Email: {\{\texttt{sisi.miao@kit.edu\}}}}\vspace{-1em}}

\maketitle

\begin{abstract}
This paper investigates the theoretical analysis of \acl{IMP} decoding for \acp{GPC} with irregular degree distributions, a generalization of \aclp{PC} that allows every code bit to be protected by a minimum of two and potentially more component codes. We derive a random hypergraph-based asymptotic performance analysis for \acp{GPC}, extending previous work that considered the case where every bit is protected by exactly two component codes. The analysis offers a new tool to guide the code design of \acp{GPC} by providing insights into the influence of degree distributions on the performance of \acp{GPC}.
\end{abstract}

\input{content_shorten}

\end{document}

%% file: macros.tex
\usepackage[dvipsnames]{xcolor}
\usepackage[cmex10]{amsmath}
\usepackage{tikz,pgfplots}
\usepackage{tikz-3dplot}
\usepackage{graphicx}
\usepackage{placeins}
\usepackage{mfirstuc}
\usepackage{mathtools}
\usepackage{wrapfig}
\usepackage{pdfpages}
\usepackage{diagbox}
\usepackage{makecell}
\usepackage{amssymb}
\usepackage{bbm}
\usepackage{bm}
\usepackage[
ruled,
vlined,
boxed, 
linesnumbered, 
commentsnumbered]{algorithm2e}
\usepackage{booktabs}%
\usetikzlibrary{backgrounds,shapes,arrows,positioning}
\usetikzlibrary{decorations.pathreplacing}
\usetikzlibrary{fit,calc} %

\usetikzlibrary{decorations.markings}
\usepackage{dblfloatfix}%
\usepackage[colorlinks=true,bookmarks=false,citecolor=black,urlcolor=black,linkcolor=black]{hyperref} %
\usepackage{cite}
\usepackage{textcomp}
\usepackage{acronym}
\let\mathcal\undefined
\DeclareMathAlphabet{\mathcal}{OMS}{cmsy}{m}{n}
\newcommand\blfootnote[1]{%
    \begingroup
    \renewcommand\thefootnote{}\footnote{#1}%
    \addtocounter{footnote}{-1}%
    \endgroup
}
\usepackage{amsthm}
\newtheorem{theorem}{Theorem}

\theoremstyle{definition}
\usepackage{enumitem}

\newtheoremstyle{definitionStyle}%
  {0pt}%
  {0pt}%
  {\normalfont}%
  {}%
  {\bfseries}%
  {.}%
  {.5em}%
  {}%
  
\theoremstyle{definitionStyle}
\newtheorem{definition}{Definition}

\newtheoremstyle{definitionStyle}%
  {0pt}%
  {0pt}%
  {\normalfont}%
  {}%
  {\bfseries}%
  {.}%
  {.5em}%
  {}%
  
\theoremstyle{definitionStyle}
\newtheorem{example}{Example}

\acrodef{LDPC}{low-density parity-check}
\acrodef{GLDPC}{generalized low-density parity-check}
\acrodef{HDD}{hard decision  decoding}
\acrodef{SDD}{soft decision decoding}
\acrodef{TPD}{turbo product decoding}
\acrodef{BSC}{binary symmetric channel}
\acrodef{iBDD}{iterative \ac{BDD}}
\acrodef{BDD}{bounded distance decoding}
\acrodef{SA-HDD}{soft-aided \ac{HDD}}
\acrodef{PC}{product code}
\acrodef{GPC}{generalized product code}
\acrodef{AD}{anchor decoding}
\acrodef{HRB}{highly reliable bit}
\acrodef{EaE}{error-and-erasure}
\acrodef{EaED}{error-and-erasure decoder}
\acrodef{SABM}{soft-aided bit marking}
\acrodef{iEaED}{iterative error-and-erasure decoding}
\acrodef{BI-AWGN}{binary input additive white Gaussian noise}
\acrodef{BCH}{Bose--Chaudhuri--Hocquenghem}
\acrodef{DRS}{dynamic reliability score}
\acrodef{DRSD}{dynamic reliability score decoder}
\acrodef{BER}{bit Error Rate}
\acrodef{SABM-SR}{SABM with scaled reliabilities}
\acrodef{NCG}{net coding gain}
\acrodef{GMD}{generalized minimal distance}
\acrodef{GMDD}{generalized minimal distance decoder}
\acrodef{RS}{Reed--Solomon}
\acrodef{DE}{decoding evolution}
\acrodef{LLR}{log-likelihood ratio}
\acrodef{BPSK}{binary phase shift keying} 
\acrodef{SNR}{signal-to-noise ratio} 
\acrodef{BEE-PC}{binary message passing based on \ac{EaE} decoding for \acp{PC}}
\acrodef{EMP}{extrinsic message passing}
\acrodef{IMP}{intrinsic message passing}
\acrodef{BCH}{Bose--Chaudhuri--Hocquenghem}
\acrodef{FER}{frame error rate}
\acrodef{BSC}{binary symmetric channel}
\acrodef{BEC}{binary erasure channel}
\acrodef{VN}{variable node}
\acrodef{CN}{check node}
\acrodef{BP}{belief propagation}
\acrodef{BP2}{binary belief propagation}
\acrodef{BP4}{quaternary belief propagation}
\acrodef{NBP}{neural \ac{BP}}
\acrodef{NBP4}{neural \ac{BP4}}
\acrodef{NBP2}{neural \ac{BP2}}
\acrodef{OBP4}{\ac{BP4} using overcomplete check matrix}
\acrodef{NOBP4}{\ac{NBP4} using an overcomplete check matrix}
\acrodef{NBP}{neural belief propagation}
\acrodef{GB}{generalized bicycle}
\acrodef{PCM}{parity check matrix}
\acrodefplural{PCM}[PCMs]{parity check matrices}
\acrodef{OSD}{ordered statistics decoding}
\acrodef{QEC}{quantum error correction}
\acrodef{QSC}{quantum stabilizer code}
\acrodef{CSS}{Calderbank–Shor–Steane}
\acrodef{NN}{neural network}
\acrodef{LLR}{log-likelihood ratio}
\acrodef{MWPM}{minimum weight perfect matching}
\acrodef{MS}{min-sum}
\acrodef{IRH}{inhomogeneous random hypergraph}
\acrodef{ESC}{extended staircase code}
\acrodef{MF}{miscorrection-free}
\acrodef{OV}{offspring vertex}
\acrodefplural{OV}[OVs]{offspring vertices}
\acrodef{RV}{random variable}
\acrodef{OH}{offspring hyperedge}
\definecolor{KITgreen}{rgb}{0,.59,.51}

\definecolor{KITpalegreen}{RGB}{130,190,60}
\colorlet{kit-maigreen100}{KITpalegreen}
\colorlet{kit-maigreen70}{KITpalegreen!70}
\colorlet{kit-maigreen50}{KITpalegreen!50}
\colorlet{kit-maigreen30}{KITpalegreen!30}
\colorlet{kit-maigreen15}{KITpalegreen!15}

\definecolor{KITblue}{rgb}{.27,.39,.66}
\definecolor{kit-blue100}{rgb}{.27,.39,.67}
\definecolor{kit-blue70}{rgb}{.49,.57,.76}
\definecolor{kit-blue50}{rgb}{.64,.69,.83}
\definecolor{kit-blue30}{rgb}{.78,.82,.9}
\definecolor{kit-blue15}{rgb}{.89,.91,.95}

\definecolor{KITyellow}{rgb}{.98,.89,0}
\definecolor{kit-yellow100}{cmyk}{0,.05,1,0}
\definecolor{kit-yellow70}{cmyk}{0,.035,.7,0}
\definecolor{kit-yellow50}{cmyk}{0,.025,.5,0}
\definecolor{kit-yellow30}{cmyk}{0,.015,.3,0}
\definecolor{kit-yellow15}{cmyk}{0,.0075,.15,0}

\definecolor{KITorange}{rgb}{.87,.60,.10}
\definecolor{kit-orange100}{cmyk}{0,.45,1,0}
\definecolor{kit-orange70}{cmyk}{0,.315,.7,0}
\definecolor{kit-orange50}{cmyk}{0,.225,.5,0}
\definecolor{kit-orange30}{cmyk}{0,.135,.3,0}
\definecolor{kit-orange15}{cmyk}{0,.0675,.15,0}

\definecolor{KITred}{rgb}{.63,.13,.13}
\definecolor{kit-red100}{cmyk}{.25,1,1,0}
\definecolor{kit-red70}{cmyk}{.175,.7,.7,0}
\definecolor{kit-red50}{cmyk}{.125,.5,.5,0}
\definecolor{kit-red30}{cmyk}{.075,.3,.3,0}
\definecolor{kit-red15}{cmyk}{.0375,.15,.15,0}

\definecolor{KITpurple}{RGB}{160,0,120}
\colorlet{kit-purple100}{KITpurple}
\colorlet{kit-purple70}{KITpurple!70}
\colorlet{kit-purple50}{KITpurple!50}
\colorlet{kit-purple30}{KITpurple!30}
\colorlet{kit-purple15}{KITpurple!15}

\definecolor{KITcyanblue}{RGB}{80,170,230}
\colorlet{kit-cyanblue100}{KITcyanblue}
\colorlet{kit-cyanblue70}{KITcyanblue!70}
\colorlet{kit-cyanblue50}{KITcyanblue!50}
\colorlet{kit-cyanblue30}{KITcyanblue!30}
\colorlet{kit-cyanblue15}{KITcyanblue!15}

\newcommand{\adv}[1]{\bm{A}^{(#1)}}
\newcommand{\adve}[1]{A^{(#1)}}
\newcommand{\dv}{r_{\textrm{max}}}

\renewcommand{\vec}[1]{\bm{#1}}
\newcommand{\liter}{^{(\ell)}}
\newcommand{\bmi}{\underline{\bm{i}}_r}
\newcommand{\cpn}{\mathsf{c}}
\newcommand{\Dc}{\mathsf{D}_{\mathsf{c}}}
\newcommand{\nrandom}{N^{\mathsf{G}}}
\usepackage{pgfplotstable}
\pgfplotsset{compat=newest} 
\usetikzlibrary{angles,quotes}
\usetikzlibrary{shapes.geometric}

\binoppenalty=\maxdimen
\relpenalty=\maxdimen

%% file: content_shorten.tex
\acresetall
\vspace{-0.4em}
\section{Introduction}
\blfootnote{This work has received funding from the European Research Council (ERC) under the European Union’s Horizon 2020 research and innovation programme (grant agreement No. 101001899) and from the 
German Federal Ministry of Education and Research (BMBF) within the project Open6GHub (grant agreement 16KISK010).}
\Acp{PC}~\cite{Elias1955} were initially introduced as a 2D array with rows and columns protected by possibly different row and column component codes. Typically, the component codes are \ac{BCH} or \ac{RS} codes, known for their large minimum distances and efficient low-complexity algebraic \ac{BDD}. Subsequently, through a spatially-coupled structure, \acp{PC} were generalized to include staircase codes~\cite{smith2011staircase}, zipper codes~\cite{sukmadji2022zipper}, braided codes~\cite{feltstrom2009braided}, and many more, resulting in enhanced decoding performance. The shared characteristic of the above-mentioned codes is that each bit is invariably protected by exactly two component codes. Utilizing the concept of \ac{GLDPC} codes~\cite{tanner1981recursive}, this leads to a Tanner graph with a regular \ac{VN} degree of $2$, the minimum value required for iterative decoding.

Apart from spatial coupling, a natural extension of \acp{PC} involves considering an increased \ac{VN} degree, either regular or irregular. A straightforward example is a 3D \ac{PC}, which offers increased error-correcting capabilities, but also leads to a substantial rise in block length and number of component code decoding steps. Using the zipper code framework, GPCs with high VN degree can be constructed with only a small decoding complexity increase. Recent works have proposed some new generalized zipper codes~\cite{sukmadji2023generalized,shehadeh2023higher}. We refer to the various generalized codes as \acp{GPC}.

Two approaches exist in the literature to study the asymptotic performance of GPCs. One studies the \ac{EMP} decoding~\cite{jian2017approaching} using \ac{GLDPC} code ensembles and density evolution. Although \ac{EMP} decoding provides an additional decoding gain compared to \ac{IMP}, it requires a modified component code decoder with increased decoding complexity and additional storage. The main advantage of the EMP-based analysis is that miscorrections can be handled rigorously. Another approach presented in~\cite{zhang2017spatially} also studies the decoding behavior of GLDPC ensembles but uses a differential equation method to study the original \ac{IMP} decoding of GPCs, where the messages passed on a certain edge are \emph{dependent} on the previous messages from the same edge, contrary to \ac{EMP}.

In this work, we use a different approach to derive a theoretical analysis of the \ac{IMP} decoding behavior of GPCs extending the previous work of Justesen~\cite{justesen2010performance} and Häger et. al.~\cite{hager2017density} where only GPCs with regular VN degree of 2 is considered, allowing for arbitrary degree distribution. The analysis is based on the theory of random graphs~\cite{erdds1959random,bollobas2009random} and the asymptotic ensembles of error patterns, where the randomness is introduced by the channel, and no code ensemble is needed. As we study a fixed interleaver defined by specific GPCs, our approach accommodates different decoding schedules and thus, is better suited for spatially-coupled codes typically decoded with a sliding window decoder. We also study an extended version of staircase codes~\cite{smith2011staircase} with an irregular degree distribution using both theoretical and simulation results.

\textbf{Notation}: Boldface letters denote vectors, matrices, and tensors, e.g., $\vec{a}$ and $\vec{A}$. Let $L^{\times r}$ be the shorthand notation for $L\times L\times \cdots \times L$ with $r$ occurrences of the number $L$. It is used in defining the shape and dimension of a tensor, e.g., $\bm{A}\in \mathbb{N}^{L^{\times r}}$. The $i$th component of vector $\vec{a}$ is denoted by $a_i$ and the elements of an $r$-dimensional tensor $\vec{A}$ are denoted by $A_{i_1,i_2,\ldots,i_r}$. The notation $\bm{A}_{:,\cdots, :, i,:,\cdots, :}$ with $i$ at the $j$th position extracts the $i$th slice of $\bm{A}$ along the $j$th dimension, yielding a tensor with dimension reduced by 1. The notation $\bm{A}_{:,i:j}$ retrieves columns $i$ to $j$ of $\bm{A}$. Matrix transpose is denoted as $\vec{A}^{\mathsf{T}}$. The set $\{1,2,\cdots,p\}$ is denoted by $[p]$ for any $p\in \mathbb{N}$. The Poisson distribution with mean $\lambda$ is denoted as $\mathsf{Po}(\lambda)$, and $\Psi_{\geq t}(\lambda)=\mathbb{P}(\mathsf{Po}(\lambda)\geq t)$. For a multiset $\underline{\bm{i}}$, the backslash notation, i.e., $\underline{\bm{i}}\backslash x$, means removing $x$ from $\underline{\bm{i}}$ once.

\section{GPCs and Random Hypergraphs}
\begin{definition}\label{def:GPC}
    A \ac{GPC} $\mathcal{C}$ is an extension of a \ac{PC} where every bit is protected by at least 2 component codes. \hfill$\Delta$
\end{definition}Tanner graphs represent the specific structure of GPCs.\begin{definition}\label{def:tannergraph}
The \emph{Tanner graph} associated with a GPC is a bipartite graph consisting of a set of $N$ \acp{CN} and a set of $M$ \acp{VN}. Every \ac{CN} $(\mathsf{c}_i)_{i\in [N]}$ represents a component code and every \ac{VN} $(\mathsf{v}_j)_{j\in [M]}$ represents a code bit. An edge between CN $\mathsf{c}_j$ and VN $\mathsf{v}_i$ exists if the VN participates in the codeword of the CN.
The set of indices of all neighboring CNs of a VN $\mathsf{v}_j$ is denoted as $\mathcal{N}(\mathsf{v}_j)$.
Define ${r_j:=|\mathcal{N}(\mathsf{v}_j)|}$ and ${\dv = \text{max}\{r_j:j\in [M]\}}$. A GPC is \emph{regular} if all VNs have the same degree and otherwise \emph{irregular}.\hfill$\Delta$
\end{definition} Our definitions extend the concept of GPC in \cite{tanner1981recursive}. Next, we propose Def.~\ref{def:connectiontensors} to encode the asymptotic structure of GPCs.

\begin{definition}\label{def:connectiontensors}
For a GPC, we divide the set of $N$ CNs of the GPC into $I$ disjoint subsets called \emph{positions} labeled from $1$ to $I$, and define a set  of \emph{connection tensors} ${\{\bm{A}\}:=\{\adv{2},\adv{3},\ldots, \adv{\dv}\}}$ such that:
\begin{enumerate}
\item[(1)] Each position $i\in[I]$ contains $N_i$ CNs and ${\sum_{i}{N_i} = N}$. Additionally, we define ${\gamma_i = N_i/N}.$
\item[(2)] For every ${r\in[r_{\text{max}}]\backslash\{1\}}$, ${\adv{r}\in \mathbb{N}_0^{I^{\times r}}}$ represents the connection of the CNs via degree $r$ VNs in the sense that for any $i_1\in[I]$ and ${i_2\leq i_3 \leq \ldots \leq i_r}$, every CN at position $i_1$ is connected to the CNs at position $i_2,i_3,\ldots, i_{r}$ via ${\adve{r}_{i_1,i_2,\ldots, i_{r}}}$ degree $r$ VNs. 
If ${i_2\leq i_3 \leq \ldots \leq i_r}$ is not fulfilled, ${\adve{r}_{i_1,i_2,\ldots, i_{r}}=0}.\hfill\Delta$
\end{enumerate}

\end{definition}

We illustrate how the connection tensors are obtained. First, the partitioning of the CNs must ensure that all CNs within the same position have the same number of degree $r$ VNs connecting to any other $r-1$ positions for any $r$. Therefore, the partition is not arbitrary, but every position should contain ``similiar'' CNs. Then, we consider the subgraph of the Tanner graph consisting of degree $r$ VNs and their neighboring CNs. The sum of the entries of ${\adv{r}_{i_1,:,\cdots, :}}$ is the degree of the CNs at position $i_1$ in this subgraph. The additional restriction, i.e., $i_2\leq i_3 \leq \ldots \leq i_r$, ensures that the number of neighboring VNs to a CN is counted exactly once.
Note that a given set of connection tensors permits non-unique Tanner graphs as it only specifies the number of connecting VNs between positions but not the exact connections between single vertices. In contrast to a code ensemble, the connections between the positions can be arbitrary but are fixed. We further clarify this in Examples~\ref{eg:pc} and~\ref{eg:pc3d}. Another example will be described later in Sec.~\ref{sec:ESC}.

\begin{figure}[t!]
\centering
\input{img/PC}
\caption{Graphical illustration of a 2D PC whose component codes are $(n_\cpn=2,k_\cpn=1)$ codes and the process of obtaining the residual graph from its Tanner graph assuming that $\mathsf{v}_1$ and $\mathsf{v}_4$ are erased.}
\label{fig:pc} 
\input{img/pc3d}
\vspace{-2ex}
\caption{Graphical illustration of a 3D PC whose component codes are $(n_\cpn=2,k_\cpn=1)$ codes and the process of obtaining the residual graph from its Tanner graph assuming that $\mathsf{v}_2$, $\mathsf{v}_6$, and $\mathsf{v}_7$ are erased.}
\label{fig:pc3d} 
\vspace{-1em}
\end{figure}
\begin{example}\label{eg:pc}
For a plain 2D PC with the same row and column component code $\mathcal{C}_{\mathsf{c}}(n_\cpn,k_\cpn)$, with an example depicted in Fig.~\ref{fig:pc}-(a), we see that every row CN is connected to the column CNs via $n$ degree $2$ VNs and vice versa. Therefore, we choose ${I=2}$ and ${N_1=N_2=n}$, corresponding to the row and column CNs defined as position 1 and 2, respectively. Then, $\{\bm{A}\} = \{\adv{2}\}$ where $\adv{2} = \left(\begin{smallmatrix}0&n\\n&0\\\end{smallmatrix}\right)$.
\end{example}

\begin{example}\label{eg:pc3d}
For a 3D cubic PC consisting of the same component code $\mathcal{C}_{\mathsf{c}}(n_\cpn,k_\cpn)$,  with an example depicted in Fig.~\ref{fig:pc3d}-(a), we notice that we can subdivide the CNs into positions 1, 2, and 3. These positions correspond to the directions along which a CN is placed, namely, along the $x$, $y$, or $z$-axis, respectively, resulting in $I=3$ and $N_1=N_2=N_3 = n^2$. We can verify that, e.g., every CN at position 1 is connected to $n$ degree 3 VNs, which then connect to $n$ CNs at position 2 and $n$ CNs at position 3. The connection tensors are $\{\bm{A}\} = \{\adv{2}, \adv{3}\}$ where $\adv{2} = \bm{0}_{3\times 3}$ and $\adv{3}$ is
	\[\adv{3}_{1,:,:} = \left(\begin{smallmatrix}0&0&0\\0&0&n\\0&0&0\\\end{smallmatrix}\right),\adv{3}_{2,:,:} = \left(\begin{smallmatrix}0&0&n\\0&0&0\\0&0&0\\\end{smallmatrix}\right),
	\adv{3}_{3,:,:} = \left(\begin{smallmatrix}0&n&0\\0&0&0\\0&0&0\\\end{smallmatrix}\right).\]
\end{example}

\begin{definition}\label{def:hypergraph}
\cite{berge1984hypergraphs} A \emph{hypergraph} $(V, E)$ is defined as a set $V$ of vertices and a set $E$ of hyperedges, which are subsets of the set of vertices, i.e., $E\subseteq\mathcal{P}(V)$. The \emph{degree} of a hyperedge $e\in E$ is its cardinality. A hypergraph is \emph{simple} if no hyperedge is a subset of another hyperedge. A hypergraph is termed \emph{$r$-uniform} if all hyperedges have cardinality $r$. A $2$-uniform hypergraph is a graph. \hfill$\Delta$
\end{definition}
A Tanner graph according to Def.~\ref{def:tannergraph} can be described as a hypergraph by
choosing the CNs as vertices, i.e., ${V=\{\mathsf{c}_{i}:i\in[N]\}}$. Then, choosing the neighboring sets of the VNs as the hyperedges of the hypergraph, i.e., ${E=\{\left\{\mathsf{c}_{i}:i\in \mathcal{N}(v_j)\right \}:{j\in [M]}\}} $. Note that all hypergraphs used in this manuscript are simple.
Next, we define residual graphs of a GPC with two examples depicted in Fig.~\ref{fig:pc} and Fig.~\ref{fig:pc3d}.
\begin{definition}
Let ${\mathcal{E}\subseteq [M]}$ be the set of indices of erased VNs. Then, a residual graph of a GPC is a hypergraph  $(V_R, E_R)$ consisting of the hyperedges associated with erased VNs and the vertices contained these hyperedges, i.e., ${E_R=\{\left\{\mathsf{c}_{i}:i\in \mathcal{N}(v_j)\right \}:{j\in \mathcal{E}}\}}$  and ${V_R=\bigcup_{e\in E_R} e}$.\hfill $\Delta$
\end{definition}
Based on \cite{soderberg2002general,bollobas2007phase,bollobas2009random,bollobas2011sparse}, we introduce the following simplified and residual graph-adapted definition.
\begin{definition}\label{def:IRHG}
Consider the vertex and hyperedge sets based on the GPC $\mathcal{C}$ defined as follows:
\begin{enumerate}
\item[(1)] Vertices: There exist $\nrandom$ vertices, each being independently assigned a \emph{type} $i\in [I]$ with probability $\gamma_i$. Let $\nrandom_i:=\nrandom\gamma_i$.
\item[(2)] Hyperedges: There exist $m_{\bmi}$ possible degree $r$ hyperedges of \emph{type} $\bmi$, where the type $\bmi:=(i_1,i_2,\ldots,i_r)$ is a multiset of the vertex types that the hyperedge contains, i.e., for $j\in[r], i_j\in [I]$. Then, a hyperedge is characterized by the tuple $(r, \bmi, m_{\bmi})$.
\end{enumerate}
A GPC-associated \emph{\ac{IRH}} $H(p,\mathcal{C})$ is a probability space where each hyperedge appears independently with probability $p$ regardless of its type.\hfill$\Delta$
\end{definition}

We first focus on a \emph{\ac{BEC}} with erasure probability $p$ and GPCs with identical component codes $\mathcal{C}_{\mathsf{c}}(n_\cpn,k_\cpn,t)$ that recover up to $t$ erasures each. The iterative decoding process entails the removal of all erasures by the component decoder $\Dc$ if it is connected to up to $t$ erasures in each iteration. 

Comparing Def.~\ref{def:connectiontensors} and Def.~\ref{def:IRHG} establishes connections between the corresponding concepts. In our analysis, every VN is erased with probability $p$ and, therefore, every hyperedge appears with probability $p$ in the \ac{IRH}. The type of vertices in the \ac{IRH} corresponds to the positions of a CN in the GPC. The parameters such as $\gamma_i$ and entries of $\adv{r}$ given by a GPC are used as constants. Then, the \ac{IRH} analysis is based on the assumption that ${\nrandom\rightarrow \infty}$ while preserving ${\nrandom_i = \nrandom\gamma_i}$ and \begin{equation}
    m_{\bmi}/\nrandom_{i_j} = \adve{r}_{i_j, \bmi\backslash i_j}\label{eq:relation}
\end{equation} for all $\bmi$ and $i_j\in \bmi$. When used as index of a tensor, the multiset $\bmi\backslash i_j$ is assumed to be sorted ascendingly.

\section{Decoding Evolution of GPCs}\label{sec:DE}

In this section, we derive Theorem~\ref{theorem:DE}, which is a mathematically tractable recursion on the bounded average of the \ac{IRH} referred to as \ac{DE}. To do so, we apply iterative peeling of the branching tree formed in a branching process~\cite[Chap. 10]{alon2016probabilistic} of the \ac{IRH}.

Consider a branching process that begins with a single root vertex at depth $\ell = 0$ and extends iteratively. Define an \ac{OH} as an incident hyperedge of the parent node at depth $\ell$ which extends to depth $\ell+1$ and define the \acp{OV} as the vertices introduced by the \acp{OH}. At every layer $\ell$, every vertex generates a random number of \acp{OH} where each \ac{OH} of type $\bmi$ introduces $r-1$ \acp{OV}. The branching process based on our \ac{IRH} is a discrete-time Markov chain $(Z_{\ell})_{\ell \geq 0}$ defined by
$Z_{0}=1$ and $Z_{\ell+1} = \sum_{j=1}^{Z_{\ell}} \xi_{\ell,j}$ representing the number of vertices at level $\ell+1$. Here, $(\xi_{\ell,j})_{\ell,j\geq 0}$ is a two-dimensional sequence of non-negative integer \acp{RV} representing the number of \acp{OV} of the ${j}$th vertex at depth $\ell$. Additionally, we define $(\zeta_{\ell,j})_{\ell,j\geq 0}$ as non-negative integer \acp{RV} representing the number of \acp{OH}. 
To determine the shape of the branching tree, we compute the distribution of $\zeta_{\ell,j}$, with which the distribution of $\xi_{\ell,j}$ is associated. Moreover, as becomes more clear in what follows, the distribution of $\zeta_{\ell,j}$ determines the \ac{DE}.

\begin{figure}
    \centering
    \input{img/branching}
    \vspace{-4ex}
    \caption{An example of one level branching process of one vertex of type $i$. Vertices are depicted as squares with its type in it. Hyperedges are depicted as a circles connecting the vertices. Possible hyperedges are shown in gray. Hyperedges are realized with probability $p$ and depicted in black. Assume that the type $i$ vertex is connected to type $j$ vertices by degree 2 hyperedges and to type $j_1$ and $j_2$ vertices by degree 3 hyperedges.}
    \vspace{-3ex}
    \label{fig:branching}
\end{figure}
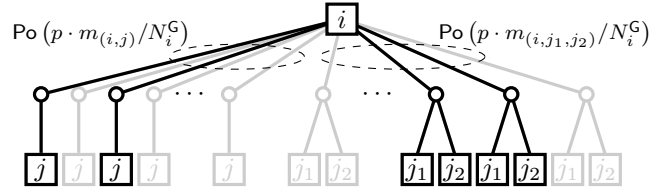
Now, we consider an arbitrary vertex of type ${i\in[I]}$ at an arbitrary depth $\ell$ as a parent vertex. Note that the number of \acp{OH} depends solely on the parent vertex type, irrespective of its depth. Then, the number of its \acp{OH} of type $\bmi$ follows $\mathsf{Po}\left(p\cdot m_{\bmi}/\nrandom_i\right)$ as $\nrandom_i\rightarrow \infty$ and the occurrence of every hyperedge is Bernoulli distributed with probability $p$. Additionally, the average number of realized \acp{OH} is $p\cdot m_{\bmi}/\nrandom_i$. Define the set of hyperedge types ${\mathcal{I}_{r,i}:= \{\bmi:i\in \bmi\}}$. Then, using the merging property of Poisson processes, the total number of \acp{OH} follows $\mathsf{Po}(\lambda_i)$ with
\begin{equation}\label{eq:lambdai}
    \lambda_i = \sum_{r=2}^{\dv}\sum_{\bmi\in \mathcal{I}_{r,i}}p\cdot m_{\bmi}/\nrandom_i.
\end{equation} An example of the branching from one vertex is illustrated in Fig.~\ref{fig:branching}. Therefore, we write that ${\zeta_{\ell,j}\sim \mathsf{Po}\left(\lambda_{\tau(Z_{j})}\right)}$ where ${\tau(Z_{j})\in[I]}$ maps $Z_{j}$ to its vertex type.  Repeating the branching process yields a branching tree.

Now, we perform an iterative batch peeling process from depth $\ell-1$ to depth $0$. In contrast to~\cite{hager2017density}, every peeling step is split into two steps. Starting at depth $\ell-1$, in the first step, given that $\Dc$ corrects up to $t$ erasures, all vertices at depth $\ell-1$ are peeled if the number of their incident hyperedges is at most $t$, including its \acp{OH} and the hyperedge extending to its parent vertex at depth $\ell-2$. 
We compute $x_{i}^{(1)} = \Psi_{\geq t}(\lambda_i)$ to be the probability of a vertex at depth $\ell-1$ and in position $i$ surviving the peeling. 
In the second step, before proceeding to depth $\ell-2$, we remove all hyperedges connecting depth $\ell-2$ to $\ell-1$ if they involve any peeled vertex. 
The interpretation is that a bit is recovered (meaning its associated hyperedge is removed) if it is recovered by any of its component decoders. Using the splitting property of Poisson processes, we know that the number of surviving \acp{OH} of a type $i$ vertex at depth $\ell-2$ is now $\mathsf{Po}\left( \lambda_i \Pi_{i_j\in \bmi\backslash{i}} x_{i_j}^{(1)}\right)$ distributed. %
The process is repeated yielding the update equation $x_{i}^{(\ell+1)} =\Psi_{\geq t}\left(\lambda_i \Pi_{i_j\in \bmi\backslash{i}} x_{i_j}^{(\ell)}\right).$
Together with \eqref{eq:relation} and \eqref{eq:lambdai}, we obtain Theorem~\ref{theorem:DE}.

\begin{theorem}\label{theorem:DE}\normalfont
     For a GPC, let ${\bm{x}\liter=\big(\begin{matrix}
         x_1\liter&x_2\liter&\cdots&x_I\liter\\
     \end{matrix}\big)}$ and $x_i\liter$ corresponds to the probability that a bit at position $i\in[I]$ is not recovered after $\ell$ decoding iterations by any of its component codes estimated by its associated IRH assuming ${\nrandom \rightarrow \infty}$. With $\bm{x}^{(0)}=\bm{1}_{1\times I}$, the evolution of $\bm{x}\liter$ is characterized as
\begin{equation}
x^{(\ell+1)}_{i} = \Psi_{\geq t} \left(p \cdot \sum_{r=2}^{\dv}\sum_{\bmi\in\mathcal{I}_{r,i}} \adve{r}_{i,\bmi\backslash i} \prod_{i_j\in \bmi\backslash{i}} x_{i_j}^{(\ell)}\right) .\label{eq:DE}
\end{equation}
\end{theorem}

\section{Decoding Schedule}
Comparing to the code ensemble based methods~\cite{jian2017approaching,zhang2017spatially}, the advantage of the random graph-based analysis is the ability to incorporate certain decoding schedules. This is especially beneficial for spatially coupled GPCs, where a windowed decoder is used instead of global decoding and also when considering the position-alternating nature of GPC decoding. For example, for the 2D PC defined in Example \ref{eg:pc}, assume that decoding starts with the row CNs at position $i=1$. Therefore, when decoding the column codes, the row codes have undergone one additional decoding step. The decoding evolution adapts by computing $x_2^{(\ell+1)}$ using the updated $x_1^{(\ell+1)}$ while computing $x_1^{(\ell+1)}$ using $x_2^{(\ell)}$.

Let $\mathcal{W}$ denote the positions that are decoded in the current window and arrange the $I$ positions such that position $i_1$ is decoded before position $i_2$ if $i_1<i_2$. The decoding evolution is modified to
\begin{equation*}\displaystyle
x^{(\ell+1)}_{i} \;\mathclap{=}\, \begin{cases}
\Psi_{\geq t} \Biggl(p  \displaystyle\sum_{r=2}^{\dv}\displaystyle\; \; \;  \;\mathclap{\sum_{\;\;\bmi\in\mathcal{I}_{r,i}}} \;\; \; \adve{r}_{i,\bmi\backslash i} \; \; \; \mathclap{\prod_{\substack{i_j\in \bmi\backslash{i},\\ i_j\geq i}}}\; \; \;  x_{i_j}^{(\ell)} \; \; \; \mathclap{\prod_{\substack{ i_j\in \bmi\backslash{i},\\  i_j< i}}}\; \; \;  x_{i_j}^{(\ell+1)}\Biggl) & \hspace{-2ex}i \in \mathcal{W}\\
x^{(\ell)}_{i} & \hspace{-2ex}i \not\in \mathcal{W}.
\end{cases}
\end{equation*}

\section{Applications of the DE}\label{sec:application}
We apply the derived DE to \acp{BSC}---where GPCs are widely used---altering the meaning of $p$ and $t$. Hereafter, $p$ denotes the cross-over probability of a \ac{BSC} and $t$ is the number of errors $\Dc$ can correct. Our conclusion on the \ac{BEC} naturally transfers to the \ac{BSC} assuming an idealized $\Dc$ correcting up to $t$ errors without any \emph{miscorrections}---instances where $\Dc$ yields a codeword $\bm{c}\in \mathcal{C}_{\mathsf{c}}$ but $\bm{c}\neq \bm{x}$, the transmitted codeword. The decoder is referred to as \ac{MF} \ac{iBDD}. The \ac{MF} assumption does not hold for most of the practically used GPCs, which often use component codes with small minimum distance and hence $t$. However, \ac{MF} decoding can be approached using, e.g., extensive component code shortening~\cite{sukmadji2023generalized,zhang2014staircase}, anchor decoding~\cite{hager2018approaching}, or soft-aided decoding algorithms~\cite{lei2019improved,miao2022JLT}, which are beyond the scope of this work.   

\vspace{-0.5ex}
\subsection{\Ac{BER} Prediction}

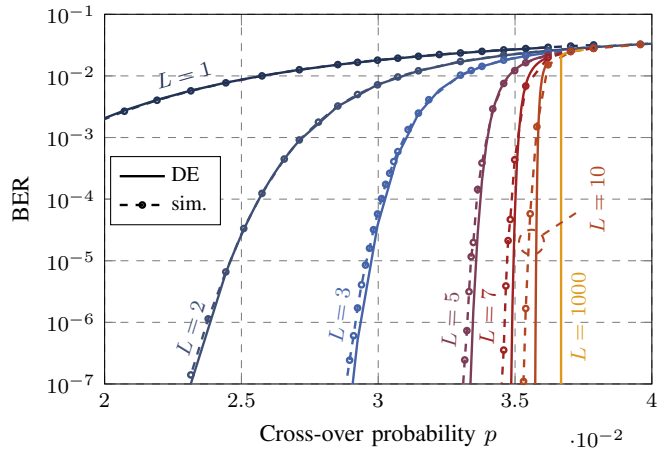
\begin{figure}[t]
	\centering
	\input{img/cubicPC}
    \vspace{-1.2em}
	\caption{DE and simulation results of the 3D PC with $(127,113,2)$ BCH component codes using $L$ iterations of \ac{MF} iBDD decoding over a \ac{BSC}.}
    \vspace{-1.1em}
	\label{fig:cubicPC_DE_sim_compare}
\end{figure}

The \ac{BER} predicted by \ac{DE} after $L$ decoding iterations is
\[p\cdot \frac{\sum_{r=2}^{\dv}\sum_{\bmi\in\mathcal{I}_{r,i}}A_{i,\bmi\backslash i}\prod_{i_j\in \bmi}x_{i_j}}{\sum_{r=2}^{\dv}\lVert \adv{r} \rVert},\]
where $\bm{x}:=\bm{x}^{(L)}$ and $\lVert \adv{r} \rVert$ is the sum of the values of the elements of $\adv{r}$. As an example, Fig.~\ref{fig:cubicPC_DE_sim_compare} compares the DE results with simulated \ac{BER} results after \ac{MF} iBDD decoding for the cubic PC with $(127,113,2)$ BCH component codes using different number of decoding iterations $L$. The theoretical and simulated results closely agree.

\begin{figure*}[t]
	\begin{minipage}[t]{0.15\textwidth}
    \centering
 \includegraphics[width=3cm]{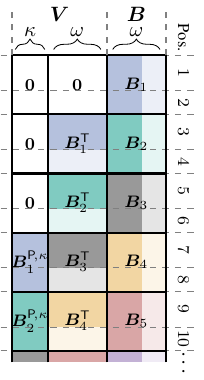}
 \vspace{-2em}
	\caption{Graphical illustration of the ESC.}
 \vspace{-1em}
	\label{fig:ESC_block}
\end{minipage}\hfill
  \begin{minipage}[t]{0.8\textwidth}
  \includegraphics[width=14cm]{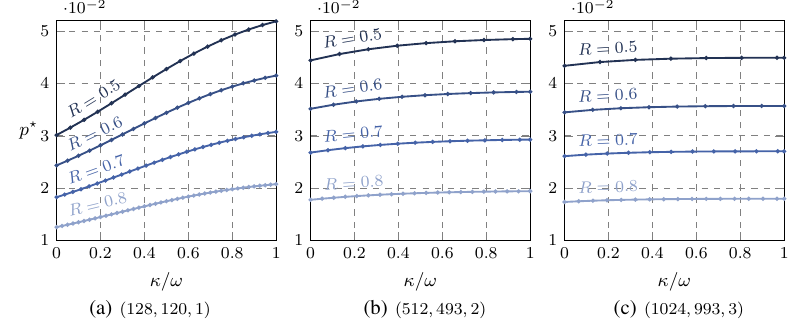}
\vspace{-0.7em}
\caption{Decoding thresholds $p^{\star}$ for ESCs of different rates $R$ based on (a) $(128,120,1)$, (b) $(512,493,2)$, and (c) $(1024,993,3)$ extended BCH component codes using DE.}
\label{fig:degree_opt}
\vspace{-1em}
 \end{minipage}
\end{figure*}

\begin{figure*}[t]
\begin{minipage}[t]{0.47\textwidth}
\centering
\input{img/nu_7_t_1_e_1}
\vspace{-1em}
\caption{Decoding results of the ESC based on $(128,120,1)$ extended BCH component codes. The ESC rate is 0.77 with $\omega=35$. We select $\kappa = 0, 18$, and 35, corresponding to staircase, half ESC, and ESC code, respectively.}
\vspace{-2ex}
\label{fig:nu_7_t_1_e_1}
\end{minipage}\hfill
\begin{minipage}[t]{0.47\textwidth}
\input{img/nu_9_t_2_e_1}
\vspace{-1em}
\caption{Decoding results of the ESC based on $(512,493,2)$ extended BCH component codes. The ESC rate is 0.8 with $\omega=95$. We select $\kappa = 0, 50$, and 95, corresponding to staircase, half ESC, and ESC code, respectively.}
\vspace{-2ex}
\label{fig:nu_9_t_2_e_1}
\end{minipage}
\end{figure*}

\subsection{Extended Staircase Code with DE}\label{sec:ESC}
We construct and inspect a simple \ac{ESC}, a generalized zipper code~\cite{sukmadji2022zipper} with flexible degree distribution, to obtain insights on the influence of the degree distribution on the decoding performance.

\subsubsection{Code Construction}
A zipper code can be described in an infinite buffer of width $n_\cpn$ where every row is protected by a component code $\mathcal{C}_\cpn(n_\cpn, k_\cpn)$. 
The buffer consists of the real part $\bm{B}$ to be transmitted over the channel and the virtual part $\bm{V}$ obtained from $\bm{B}$ via a specific mapping. 
To describe an \ac{ESC}, we divide every $\omega$ rows of the buffer into one spatial position indicated by a subscript $i\in \mathbb{N}$, e.g., $\bm{B}_i$ and $\bm{V}_i$. Addtionally, define a configurable parameter $\kappa$.

Upon encoding, one block $\bm{B}_i \in \mathbb{F}_2^{\omega \times \omega}$ is generated at a time where $\left(\bm{B}_i\right)_{:,1:(k_{\cpn}-\omega -\kappa)}$ are new information bits and $\left(\bm{B}_i\right)_{:,(k_\cpn-\omega -\kappa+1):\omega}$ are parity check bits obtained by encoding the rows at spatial position $i$. 
To do so, we partition $\bm{V}_{i} = \left(\bm{V}_i^{(1)}\; \bm{V}_i^{(2)}\right) \in \mathbb{F}_2^{\omega\times (\kappa+\omega)}$ for all $i$ and define the mapping as $\bm{V}_{i+1}^{(2)} = \bm{B}_i^{\mathsf{T}} \in \mathbb{F}_2^{\omega \times \omega}$ and $\bm{V}_{i+3}^{(1)} = \left(\bm{B}_i\right)^{\mathsf{P},\kappa} \in \mathbb{F}_2^{\omega \times \kappa}$
where the operator $\left(\bm{X}\right)^{\mathsf{P},\kappa}$ is a cyclic permutation of the first $\kappa$ columns of $\bm{X}$, 
i.e., $\left(\bm{X}^{\mathsf{P},\kappa}\right)_{(((i+j-1)\;\text{mod}\;{\omega})+1),j} = \bm{X}_{i,j}$ for $j\in [\kappa]$. The mappings are defined to fulfill the general rule of thumb for mitigating high error floors by avoiding any two bits appearing in the same row more than once. A more extensive study of the mappings is presented in~\cite{shehadeh2023higher}.

To match a design code rate $R$, we choose a real buffer width $\omega$ such that $\frac{\omega-(n_\cpn-k_\cpn)}{\omega} = R$, regardless of the degree distribution. To ensure an integer number of bits, $\omega$ is rounded to the nearest integer, causing a negligible deviation from $R$. We can then construct ESCs with different degree distributions for the same code rate $R$. We use a $(2^{\nu},2^{\nu}-t\nu,t)$ extended BCH code and shorten it such that $n_\cpn = \kappa+2\omega$. To have a valid ESC, $\omega>t\nu$, $2\omega  + \kappa \leq 2^{\nu}$ and $0\leq \kappa \leq \omega$ must hold. When $\kappa=0$, the code is a staircase code. When $\kappa=\omega $, the code is a regular generalized zipper code with VN degree $3$. The structure of the ESC is depicted in Fig.~\ref{fig:ESC_block} where dark and light-colored parts consists of degree 3 and degree 2 VNs, respectively.
	
We proceed to the connection tensors of the ESC. First, the positions are obtained by partitioning every spatial position $i$ into two positions $i_u$ and $i_l$ containing the first $\kappa$ rows and the last $\omega-\kappa$ rows, respectively. Then, for an ESC with $I'$ spatial positions, we have $I=2I'$, $N_{2i-1} = \kappa$, and $N_{2i} = \omega-\kappa$ for all $i\in [I']$. Naturally, $\bm{A} = \{\adv{2}, \adv{3}\}$ as $r_{\text{max}}=3$. We first compute the entries of $\adv{2}$. For odd $i$ and ${i\in [I-3]}$, we have ${\adve{2}_{i,i+3}=\omega-\kappa}$ and ${\adve{2}_{i+3,i}=\kappa}$. For even $i$ and ${i\in [I-2]}$, we have ${\adve{2}_{i,i+2}=\adve{2}_{i+2,i}=\omega-\kappa}$. We then compute the entries of $\adv{3}$ using an approximation that the bits in the upper and lower parts of $\bm{V}_{i}^{(1)}$ are treated as identical for all $i$.\footnote{Such approximation is often needed when handling GPCs where the interleaving of the bits are relatively complex, e.g., the Chevron codes~\cite{sukmadji2023generalized}. However, a good alignment of the DE and simulated results is still observed.} Then, for odd $i$ and $i\in [I-6]$, we have ${\adve{3}_{i,i+2,i+6}=\adve{3}_{i+2,i,i+6}=\adve{3}_{i+6,i,i+2}=\kappa}$. For even $i$ and $i\in [I-6]$, we have ${\adve{3}_{i,i+1,i+6}=\adve{2}_{i+6,i,i+1}=\kappa}$ and ${\adve{3}_{i+1,i,i+6}=\omega-\kappa}$. All other entries of $\adv{2}$ and $\adv{3}$ are $0$.

\subsubsection{Results}

The impact of the degree distribution on the decoding performance of \acp{ESC} is examined using a fixed window length of $7$ and $8$ decoding iterations. The decoding threshold $p^{\star}$ denotes the minimum $p$ value for which the DE-estimated \ac{BER} falls below $10^{-4}$. Consistently, an increase in the fraction of degree 3 \acp{VN} leads to an improvement in the decoding threshold, with greater improvements observed as $t$ decreases. Notably, for the rarely used $t=1$ codes, introducing high-degree VNs results in the most significant improvement. Figure~\ref{fig:degree_opt} depict the results of $p^\star$ for three ESC families based on the $(128, 120, t=1)$, $(512, 493, t=2)$, and $(1024, 993, t=3)$ extended BCH codes. Similar trends are observed for other component codes with various lengths.

Based on the observations of decoding thresholds, we take a closer look on two sets of exemplary codes based on $(128,120,1)$ and $(512,493,2)$ extended BCH codes using MF iBDD, as depicted in Fig.~\ref{fig:nu_7_t_1_e_1} and Fig.~\ref{fig:nu_9_t_2_e_1}, respectively. The post-decoding \ac{BER} improvements achieved by leveraging high-degree VNs align with the DE predictions. Particularly, for the $t=1$ code, a substantial decoding gain is observed, while for the $t=2$ code, introducing approximately 50\% degree 3 VNs leads to noticeable improvements. Furthermore, beyond the decoding threshold improvement, we heuristically deduce that high-degree ESCs exhibit a lower error floor compared to staircase codes. This aligns with observations in~\cite{sukmadji2023generalized} and \cite{shehadeh2023higher}, although a systematic error floor analysis of irregular GPC has not been conducted. Note that in practice, MF iBDD can be approached using the aforementioned methods ~\cite{sukmadji2023generalized,zhang2014staircase,hager2018approaching,lei2019improved,miao2022JLT}.

\vspace{-0.5em}
\section{Conclusion}
\vspace{-0.5em}
We analyze the asymptotic decoding behavior of GPCs with irregular degree distribution through the derived decoding evolution. As an example, we examine a simple extension of the staircase code. Future work includes, e.g., extending the analysis to a mixture of different component codes.

\ifCLASSOPTIONcaptionsoff
  \newpage
\fi

\IEEEtriggeratref{12}
\bibliographystyle{IEEEtran}

\bibliography{IEEEabrv,literature}

%% file: img/PC.tex
\begin{tabular}{@{}c@{\quad}c@{}@{\quad}c@{}}
\begin{tikzpicture}[scale=0.7]
      \fill[color=kit-blue100] (0,1) rectangle (1,2); %
      \fill[color=kit-blue70] (1,1) rectangle (2,2); %
      \fill[color=kit-blue30] (0,0) rectangle (1,1); %
      \fill[color=kit-blue15] (1,0) rectangle (2,1); %

        \node[draw = none] at (0.5,1.5) {$\mathsf{v}_1$};
        \node[draw = none] at (1.5,1.5) {$\mathsf{v}_2$};
        \node[draw = none] at (0.5,0.5) {$\mathsf{v}_3$};
        \node[draw = none] at (1.5,0.5) {$\mathsf{v}_4$};

        \node[draw = none] at (2.5,1.5) {$\mathsf{c}_1$};
        \node[draw = none] at (2.5,0.5) {$\mathsf{c}_2$};
        \node[draw = none] at (0.5,-0.3) {$\mathsf{c}_3$};
        \node[draw = none] at (1.5,-0.3) {$\mathsf{c}_4$};

  \draw[step=1cm,black, very thick] (0,0) grid (2,2);
  
\end{tikzpicture} & \begin{tikzpicture}
  \draw [fill=kit-blue100, kit-blue100] (0+0.125,1.0) circle (0.1);
  \draw [fill=kit-blue70, kit-blue70] (0.5+0.125,1.0) circle (0.1);
  \draw [fill=kit-blue30, kit-blue30] (1+0.125,1.0) circle (0.1);
  \draw [fill=kit-blue15, kit-blue15] (1.5+0.125,1.0) circle (0.1);

  \draw[kit-blue100, very thick] (0+0.125,1.0) -- (0,0);
  \draw[kit-blue100, very thick] (0+0.125,1.0) -- (1.25,0);

  \draw[kit-blue70, very thick] (0.5+0.125,1.0) -- (0,0);
  \draw[kit-blue70, very thick] (0.5+0.125,1.0) -- (1.75,0);

  \draw[kit-blue30, very thick] (1+0.125,1.0) -- (0.5,0);
  \draw[kit-blue30, very thick] (1+0.125,1.0) -- (1.25,0);

  \draw[kit-blue15, very thick] (1.5+0.125,1.0) -- (0.5,0);
  \draw[kit-blue15, very thick] (1.5+0.125,1.0) -- (1.75,0);

  \foreach \x in {0,1} {
    \draw [fill=black] (\x/2-0.125,-0.125) rectangle ++(0.25,0.25);
  }

    \foreach \x in {2,3} {
    \draw [fill=black] (\x/2+0.125,-0.125) rectangle ++(0.25,0.25);
  }
  
  \foreach \x in {1,2} { 
    \node[draw = none] at (\x/2-0.5,-0.3) {$\mathsf{c}_{\x}$};
    \node[draw = none] at (\x/2-0.5,1.3) {$\mathsf{v}_{\x}$};
  }

    \foreach \x in {3,4} { 
    \node[draw = none] at (\x/2-0.5+0.25,-0.3) {$\mathsf{c}_{\x}$};
    \node[draw = none] at (\x/2-0.5,1.3) {$\mathsf{v}_{\x}$};
  }
  
\end{tikzpicture}& \begin{tikzpicture}

  \draw[kit-blue100, very thick] (0+0.125,1.0) -- (0,0);
  \draw[kit-blue100, very thick] (0+0.125,1.0) -- (1.25,0);

  \draw[kit-blue15, very thick] (1.5+0.125,1.0) -- (0.5,0);
  \draw[kit-blue15, very thick] (1.5+0.125,1.0) -- (1.75,0);

  \foreach \x in {0,1} {
    \draw [fill=black] (\x/2-0.125,-0.125) rectangle ++(0.25,0.25);
  }

    \foreach \x in {2,3} {
    \draw [fill=black] (\x/2+0.125,-0.125) rectangle ++(0.25,0.25);
  }
  
  \foreach \x in {1} { 
    \node[draw = none] at (\x/2-0.5,-0.3) {$\mathsf{c}_{\x}$};
    \node[draw = none] at (\x/2-0.5,1.3) {$\mathsf{v}_{\x}$};
  }

    \foreach \x in {4} { 
    \node[draw = none] at (\x/2-0.5+0.25,-0.3) {$\mathsf{c}_{\x}$};
    \node[draw = none] at (\x/2-0.5,1.3) {$\mathsf{v}_{\x}$};
  }
  \end{tikzpicture}
\\
{\footnotesize (a) PC} & {\footnotesize (b) Tanner graph} & {\footnotesize (c) Residual graph}
\end{tabular}

%% file: img/pc3d.tex
\begin{tabular}{@{}c@{}c@{}}
\tdplotsetmaincoords{60}{125}
\begin{tikzpicture}
		[cube/.style={very thick,black},
			grid/.style={very thin,gray},
			axis/.style={->,black!50,thick,dashed},
   scale = 0.4]

	\draw[axis] (0,0,0) -- (2.5,0,0) node[anchor=west]{$y$};
	\draw[axis] (0,0,0) -- (0,2.5,0) node[anchor=west]{$z$};
	\draw[axis] (0,0,0) -- (0,0,3) node[anchor=west]{$x$};
 
        \draw[cube] (0,2,0) -- (2,2,0) -- (2,0,0);
       
	\draw[cube] (0,0,2) -- (0,2,2) -- (2,2,2) -- (2,0,2) -- cycle;
 
	\draw[cube] (0,2,0) -- (0,2,2);
	\draw[cube] (2,0,0) -- (2,0,2);
	\draw[cube] (2,2,0) -- (2,2,2);

        \draw[cube] (2,1,0) -- (2,1,2) -- (0,1,2);
        \draw[cube] (2,0,1) -- (2,2,1) -- (0,2,1);
        \draw[cube] (1,0,2) -- (1,2,2) -- (1,2,0);
  \end{tikzpicture}&
  \begin{tikzpicture}
  \foreach \x in {1,2,3,4} {
    \node [fill=black, rectangle] at (\x/2,0) {};
    \node[draw = none] at (\x/2,-0.3) {$\mathsf{c}_{\x}$};
    \node [draw=none] (c\x) at (\x/2,0) {};
  }

  \foreach \x in {5,6,7,8} {
    \node [fill=black, rectangle] at (\x/2+0.25,0) {};
    \node[draw = none] at (\x/2+0.25,-0.3) {$\mathsf{c}_{\x}$};
    \node [draw=none] (c\x) at (\x/2+0.25,0) {};
  }

  \foreach \x in {9,10,11,12} {
    \node [fill=black, rectangle] at (\x/2+0.5,0) {};
    \node[draw = none] at (\x/2+0.5,-0.3) {$\mathsf{c}_{\x}$};
    \node [draw=none, outer sep = 0] (c\x) at (\x/2+0.5,0.02) {};
  }

  \foreach \x in {1,2,3,4,5,6,7,8} { 
    \node[draw = none]  at (\x/4*3,1.3) {$\mathsf{v}_{\x}$};
    \node [draw=none] (v\x) at (\x/4*3,1) {};
  }

  \draw[kit-blue100, thick] (c1) -- (1/4*3,1) -- (c5) -- (1/4*3,1) -- (c9);
  \draw[kit-blue100!90, thick] (c1) -- (2/4*3,1) -- (c6) --(2/4*3,1) -- (c10);
  \draw[kit-blue100!80, thick] (c2) -- (3/4*3,1) -- (c5) -- (3/4*3,1) -- (c11);
  \draw[kit-blue100!70, thick] (c2) -- (3,1) -- (c6) -- (3,1) -- (c12);

  \draw[kit-blue100!60, thick] (c3) -- (5/4*3,1) -- (c7) -- (5/4*3,1) -- (c9);
  \draw[kit-blue100!50, thick] (c3) -- (6/4*3,1) -- (c8) -- (6/4*3,1) -- (c10);
  \draw[kit-blue100!40, thick] (c4) -- (7/4*3,1) -- (c7) -- (7/4*3,1) -- (c11);
  \draw[kit-blue100!30, thick] (c4) -- (8/4*3,1) -- (c8) -- (8/4*3,1) -- (c12);
  \draw[decorate, decoration={brace, amplitude=5pt,mirror}] (0.25,-0.45) -- (2.25,-0.45);
  \draw[decorate, decoration={brace, amplitude=5pt,mirror}] (2.5,-0.45) -- (4.5,-0.45);
  \draw[decorate, decoration={brace, amplitude=5pt,mirror}] (4.75,-0.45) -- (6.75,-0.45);
  \node[draw=none, at={(1.25,-0.8)}]  {$x$-CNs};
  \node[draw=none, at={(3.5,-0.8)}]  {$y$-CNs};
  \node[draw=none, at={(5.75,-0.8)}]  {$z$-CNs};

  \foreach \x in {100,90,80,70,60,50,40,30} { 
    \draw [draw = none, fill=kit-blue100!\x] (8.25-\x/40*3,1.0) circle (0.1);
  }
  
\end{tikzpicture}\\[-6pt]
{\footnotesize (a) 3D PC} & {\footnotesize (b) Tanner graph}
\\
{}&{\begin{tikzpicture}
  \foreach \x in {1,2,3,4} {
    \node [fill=black, rectangle] at (\x/2,0) {};
    \node[draw = none] at (\x/2,-0.3) {$\mathsf{c}_{\x}$};
    \node [draw=none] (c\x) at (\x/2,0) {};
  }

  \foreach \x in {5,6,7,8} {
    \node [fill=black, rectangle] at (\x/2+0.25,0) {};
    \node[draw = none] at (\x/2+0.25,-0.3) {$\mathsf{c}_{\x}$};
    \node [draw=none] (c\x) at (\x/2+0.25,0) {};
  }

  \foreach \x in {9,10,11,12} {
    \node [fill=black, rectangle] at (\x/2+0.5,0) {};
    \node[draw = none] at (\x/2+0.5,-0.3) {$\mathsf{c}_{\x}$};
    \node [draw=none] (c\x) at (\x/2+0.5,0) {};
  }

  \foreach \x in {2,6,7} { 
    \node[draw = none]  at (\x/4*3,1.3) {$\mathsf{v}_{\x}$};
    \node [draw=none] (v\x) at (\x/4*3,1.0) {};
  }

  \draw[kit-blue100!90, thick] (c1) -- (2/4*3,1.0) -- (c6);
  \draw[kit-blue100!90, thick] -- (2/4*3+0.01,1.0) -- (c10);
\draw[kit-blue100!50, thick] (c3) -- (6/4*3,1.0) -- (c8);
\draw[kit-blue100!50, thick] (6/4*3,1.01) -- (c10);
  \draw[kit-blue100!30, thick] (c4) -- (7/4*3,1.0) -- (c7);
  \draw[kit-blue100!30, thick] (7/4*3+0.04,1.035) -- (c11);
  
  \node[ellipse,draw,dashed, anchor = center,at={(v2)},minimum width = 0.7cm, 
    minimum height = 0.2cm] (e2)  {};
    \node[ellipse,draw,dashed, anchor = center,at={(v6),yshift=-0.2cm},minimum width = 0.7cm, 
    minimum height = 0.2cm] (e3)  {};

    \node[ellipse,draw,dashed, anchor = center,at={(v7),yshift=-0.2cm},minimum width = 0.7cm, 
    minimum height = 0.2cm] (e4)  {};
    
    \node[draw=none,at={(3.2,1.2)}] (he) {hyperedge};
    \draw[black,dashed] (e2) -- (2.4,1.2);
    \draw[black,dashed] (e3) -- (3.95,1.1);
    \draw[black,dashed] (e4) -- (3.95,1.2);
    
\end{tikzpicture}}\\[-6pt]
{} & {\footnotesize (c) Residual graph}
\end{tabular}

%% file: img/branching.tex
\begin{tikzpicture}

\node [draw = none] (root) at (4,4) {};

\foreach \x in {0,1,2,3,5} {
\node[draw, black!20, very thick, inner sep=0pt, minimum size=0.4cm, name=c\x] at (\x/2,2){};  
\node [draw=none] at (\x/2,2) {\textcolor{black!20}{$j$}};
\node[draw=none] (v\x) at (\x/2,3) {};
\draw[very thick,black!20] (root) -- (\x/2,3);
\draw[very thick,black!20] (c\x) -- (\x/2,3);
\node[circle, draw, black!20, fill=white, very thick, outer sep=0pt,inner sep=0pt, minimum size=5pt]at (\x/2,3) {};
}
\node [draw=none, very thick] at (2,3) {$\cdots$};
\node [draw=none, very thick] at (4.5,3) {$\cdots$};

\foreach \x in {7,8,10,11,12,13,14,15} {
\node[draw, black!20, very thick, inner sep=0pt, minimum size=0.4cm, name=c\x] at (\x/2,2){};  
}
\foreach \x in {7,10,12,14} {
 \node [very thick,draw=none] at (\x/2,2) {\textcolor{black!20}{$j_1$}};
 \node [very thick,draw=none] at (\x/2+0.5,2) {\textcolor{black!20}{$j_2$}};
}

\foreach \x in {7,10,12,14} {
\draw[very thick,black!20] (root) -- (\x/2+0.25,3);
\draw[very thick,black!20] (\x/2,2.2)--(\x/2+0.25,3);
\pgfmathsetmacro\y{\x + 1} 
\draw[very thick,black!20] (\x/2+0.5,2.2) -- (\x/2+0.25,3);
\node[circle, draw, black!20, fill = white, very thick, outer sep=0pt, inner sep=0pt,minimum size=5pt, name=v\x] at (\x/2+0.25,3) {};
}

\foreach \x in {0,2} {
\node[draw, black, very thick, inner sep=0pt, minimum size=0.4cm] at (\x/2,2){};  
\node [draw=none] at (\x/2,2) {$j$};
\draw[very thick,black] (root) -- (\x/2,3);
\draw[very thick,black] (c\x) -- (\x/2,3);
\node[circle, draw=black, very thick, outer sep=0pt, inner sep=0pt, minimum size=5pt, fill = white]at (\x/2,3) {};
}

\foreach \x in {10,11,12,13} {
\node[draw, black, very thick, outer sep=0pt, minimum size=0.4cm] at (\x/2,2){};  
}
\foreach \x in {10,12} {
 \node [very thick,draw=none] at (\x/2,2) {\textcolor{black}{$j_1$}};
 \node [very thick,draw=none] at (\x/2+0.5,2) {\textcolor{black}{$j_2$}};
}

\foreach \x in {10,12} {
\node[circle, draw=black, very thick, outer sep=0pt, inner sep=0pt, minimum size=5pt] at (\x/2+0.25,3) {};
\draw[very thick,black] (root) -- (v\x);
\draw[very thick,black] (\x/2,2.2)--(v\x);
\draw[very thick,black] (\x/2+0.5,2.2) -- (v\x);
} 
\draw[very thick,black,fill=white] (3.8,3.8) rectangle ++(0.4,0.4);

\node[ellipse,draw,dashed,anchor = center,at={(2.6,3.5)},minimum width = 1.8cm, minimum height = 0.2] (e1)  {};
\node[ellipse,draw,dashed,anchor = center,at={(4.8,3.5)},minimum width = 2.2cm, minimum height = 0.2] (e2)  {};
\node [draw=none] at (0.8,3.8) {\footnotesize $\mathsf{Po}\left(p\cdot m_{(i,j)}/\nrandom_i\right)$};
\node [draw=none] at (6.7,3.8) {\footnotesize $\mathsf{Po}\left(p\cdot m_{(i,j_1,j_2)}/{\nrandom_i}\right)$};

\node [draw = none] at (4,4) {$i$};

\end{tikzpicture}

%% file: img/cubicPC.tex
\begin{tikzpicture}
\pgfplotsset{grid style={dashed, gray}}
\pgfplotsset{every tick label/.append style={font=\footnotesize}}
\begin{axis}[%
width=\columnwidth,
height=6.5cm,
 xmin=0.02,
 xmax=0.04,
ymode=log,
ymin=1e-7,
ymax=0.1,
yminorticks,
axis background/.style={fill=white, mark size=1pt},
xmajorgrids,
ymajorgrids,
yminorgrids,
x tick label style={
                /pgf/number format/.cd,
                    fixed relative,
                    precision=2,
                    zerofill,
                /tikz/.cd,
            },
ytick={0.1,0.01,0.001,1e-4,1e-5,1e-6,1e-7,1e-8},
ylabel={BER},
xlabel={Cross-over probability $p$},
label style={font=\small},
legend cell align={left},
legend style={anchor =north west, at={(axis cs: 0.0202,5e-4)},draw=black, fill opacity=1, text opacity = 1,legend columns=1, row sep = 0pt,font=\footnotesize}
]

\addplot [color=black, line width=0.9pt, mark=none]table[x=delta, y=BER, col sep=semicolon,row sep=crcr] {
EbNo;EsNo;delta;decodedFrame;FrameErr;FER;BER\\
2.83;0;0;0;0;0;1e-100\\
};
\addlegendentry{DE}

\addplot [color=black, dashed, line width=0.9pt, mark=o, mark options={solid,fill=white, mark size=1pt}]table[x=delta, y=BER, col sep=semicolon,row sep=crcr] {
EbNo;EsNo;delta;decodedFrame;FrameErr;FER;BER\\
2.83;-0;0;0;0;0;1e-100\\
};
\addlegendentry{sim.}

\addplot [color=black!50!KITblue,dashed, line width=0.9pt, mark=o, mark options={solid,fill=white, mark size=1pt}]table[x=delta, y=BER, col sep=semicolon,row sep=crcr] {
EbNo;EsNo;delta;decodedFrame;FrameErr;FER;BER\\
3.5;1.97824;0.0378718;157;157;1;0.0317952\\
3.55;2.02824;0.0370344;156;156;1;0.0304515\\
3.6;2.07824;0.0362076;157;157;1;0.0291238\\
3.65;2.12824;0.0353912;157;157;1;0.0277795\\
3.7;2.17824;0.0345854;155;155;1;0.0264242\\
3.75;2.22824;0.0337903;156;156;1;0.0250149\\
3.8;2.27824;0.0330058;155;155;1;0.0235943\\
3.85;2.32824;0.0322319;221;221;1;0.0222139\\
3.9;2.37824;0.0314688;212;212;1;0.0208161\\
3.95;2.42824;0.0307164;209;209;1;0.0193892\\
4;2.47824;0.0299747;209;209;1;0.0180064\\
4.05;2.52824;0.0292438;213;213;1;0.0165995\\
4.1;2.57824;0.0285236;211;211;1;0.0152626\\
4.2;2.67824;0.0271157;212;212;1;0.012587\\
4.3;2.77824;0.0257509;211;211;1;0.0100072\\
4.4;2.87824;0.0244294;215;215;1;0.00772151\\
4.5;2.97824;0.0231509;221;221;1;0.00572268\\
4.6;3.07824;0.0219154;220;220;1;0.0040273\\
4.7;3.17824;0.0207227;216;216;1;0.00267209\\
4.8;3.27824;0.0195726;221;221;1;0.0016761\\
}node [pos=0.6,anchor=south,font=\footnotesize,sloped,xshift = 1] {$L=1$};

\addplot [color=black!50!KITblue,  line width=0.9pt, mark=none]table[x=delta, y=BER, col sep=semicolon,row sep=crcr] {
EbNo;EsNo;delta;decodedFrame;FrameErr;FER;BER\\
3.6;2.0782;0.036208;0;0;0;0.02896\\
3.7;2.1782;0.034585;0;0;0;0.026255\\
3.8;2.2782;0.033006;0;0;0;0.023503\\
3.9;2.3782;0.031469;0;0;0;0.020724\\
4;2.4782;0.029975;0;0;0;0.017944\\
4.1;2.5782;0.028524;0;0;0;0.015204\\
4.2;2.6782;0.027116;0;0;0;0.012555\\
4.3;2.7782;0.025751;0;0;0;0.010058\\
4.4;2.8782;0.024429;0;0;0;0.0077768\\
4.5;2.9782;0.023151;0;0;0;0.0057715\\
4.6;3.0782;0.021915;0;0;0;0.0040877\\
4.7;3.1782;0.020723;0;0;0;0.0027467\\
4.8;3.2782;0.019573;0;0;0;0.0017411\\
};

 \addplot [color=black!40!KITblue!90,  line width=0.9pt, mark = none]
table[x=delta, y=BER, col sep=semicolon,row sep=crcr]{%
EbNo;EsNo;delta;decodedFrame;FrameErr;FER;BER\\
3;1.4782;0.046811;0;0;0;0.0437\\
3.1;1.5782;0.044942;0;0;0;0.041139\\
3.2;1.6782;0.043114;0;0;0;0.038462\\
3.3;1.7782;0.041325;0;0;0;0.035631\\
3.4;1.8782;0.039578;0;0;0;0.032593\\
3.5;1.9782;0.037872;0;0;0;0.029284\\
3.6;2.0782;0.036208;0;0;0;0.025625\\
3.7;2.1782;0.034585;0;0;0;0.021539\\
3.8;2.2782;0.033006;0;0;0;0.016987\\
3.9;2.3782;0.031469;0;0;0;0.012071\\
4;2.4782;0.029975;0;0;0;0.0072042\\
4.1;2.5782;0.028524;0;0;0;0.0032109\\
4.2;2.6782;0.027116;0;0;0;0.00089222\\
4.21;2.6882;0.026977;0;0;0;0.00075812\\
4.22;2.6982;0.026839;0;0;0;0.00063946\\
4.23;2.7082;0.026702;0;0;0;0.00053528\\
4.24;2.7182;0.026565;0;0;0;0.00044456\\
4.25;2.7282;0.026428;0;0;0;0.00036623\\
4.26;2.7382;0.026292;0;0;0;0.00029917\\
4.27;2.7482;0.026156;0;0;0;0.00024228\\
4.28;2.7582;0.02602;0;0;0;0.00019447\\
4.29;2.7682;0.025885;0;0;0;0.00015467\\
4.3;2.7782;0.025751;0;0;0;0.00012186\\
4.31;2.7882;0.025617;0;0;0;9.508e-05\\
4.32;2.7982;0.025483;0;0;0;7.3455e-05\\
4.33;2.8082;0.02535;0;0;0;5.6174e-05\\
4.34;2.8182;0.025217;0;0;0;4.2514e-05\\
4.35;2.8282;0.025085;0;0;0;3.1836e-05\\
4.36;2.8382;0.024953;0;0;0;2.3582e-05\\
4.37;2.8482;0.024821;0;0;0;1.7276e-05\\
4.38;2.8582;0.02469;0;0;0;1.2514e-05\\
4.39;2.8682;0.02456;0;0;0;8.9606e-06\\
4.4;2.8782;0.024429;0;0;0;6.3418e-06\\
4.5;2.9782;0.023151;0;0;0;1.009e-07\\
4.6;3.0782;0.021915;0;0;0;4.141e-10\\
}node [pos=0.6,anchor=south,font=\footnotesize,sloped] {$L=2$};

\addplot [color=black!40!KITblue!90, dashed,line width=0.9pt, mark=o, mark options={solid,fill=white, mark size=1pt}]table[x=delta, y=BER, col sep=semicolon,row sep=crcr] {
EbNo;EsNo;delta;decodedFrame;FrameErr;FER;BER\\
3;1.47824;0.0468113;156;156;1;0.0439282\\
3.1;1.57824;0.0449425;157;157;1;0.0413719\\
3.2;1.67824;0.0431136;156;156;1;0.0386525\\
3.3;1.77824;0.0413253;157;157;1;0.0358658\\
3.4;1.87824;0.0395778;152;152;1;0.0328231\\
3.5;1.97824;0.0378718;156;156;1;0.0294736\\
3.6;2.07824;0.0362076;157;157;1;0.0258646\\
3.7;2.17824;0.0345854;156;156;1;0.0217201\\
3.8;2.27824;0.0330058;157;157;1;0.0171484\\
3.9;2.37824;0.0314688;157;157;1;0.0121073\\
3.95;2.42824;0.0307164;155;155;1;0.0096105\\
4;2.47824;0.0299747;157;157;1;0.00716311\\
4.05;2.52824;0.0292438;154;154;1;0.00499055\\
4.1;2.57824;0.0285236;155;155;1;0.0032458\\
4.15;2.62824;0.0278143;156;156;1;0.00177638\\
4.2;2.67824;0.0271157;157;157;1;0.00091242\\
4.24;2.71824;0.0265646;154;154;1;0.000447319\\
4.3;2.77824;0.0257509;154;154;1;0.000123943\\
4.35;2.82824;0.0250848;157;157;1;3.33944e-05\\
4.4;2.87824;0.0244294;157;151;0.961783;6.58177e-06\\
4.45;2.92824;0.0237847;191;80;0.418848;1.13936e-06\\
4.5;2.97824;0.0231509;578;39;0.067474;1.40289e-07\\
};

\addplot [color=KITblue, dashed,line width=0.9pt, mark=o, mark options={solid,fill=white, mark size=1pt}]table[x=delta, y=BER, col sep=semicolon,row sep=crcr] {
EbNo;EsNo;delta;decodedFrame;FrameErr;FER;BER\\
3.5;1.97824;0.0378718;155;155;1;0.0286518\\
3.55;2.02824;0.0370344;157;157;1;0.0265418\\
3.6;2.07824;0.0362076;156;156;1;0.0240637\\
3.65;2.12824;0.0353912;155;155;1;0.0212651\\
3.7;2.17824;0.0345854;154;154;1;0.0180238\\
3.75;2.22824;0.0337903;157;157;1;0.0142891\\
3.775;2.25324;0.0333967;157;157;1;0.0122195\\
3.8;2.27824;0.0330058;154;154;1;0.0102758\\
3.875;2.35324;0.031849;215;215;1;0.00410644\\
3.9;2.37824;0.0314688;215;215;1;0.00248663\\
3.925;2.40324;0.0310912;221;221;1;0.00134569\\
3.95;2.42824;0.0307164;221;221;1;0.000590121\\
3.96;2.43824;0.0305672;218;218;1;0.000407288\\
3.97;2.44824;0.0304184;221;221;1;0.00026297\\
3.98;2.45824;0.0302701;221;221;1;0.000173824\\
3.99;2.46824;0.0301222;221;221;1;0.000102038\\
4;2.47824;0.0299747;221;221;1;5.74824e-05\\
4.01;2.48824;0.0298276;221;221;1;3.19117e-05\\
4.02;2.49824;0.029681;221;218;0.986425;1.59276e-05\\
4.03;2.50824;0.0295348;223;204;0.914798;8.50618e-06\\
4.04;2.51824;0.0293891;236;175;0.741525;4.06726e-06\\
4.05;2.52824;0.0292438;262;129;0.492366;1.71146e-06\\
4.06;2.53824;0.0290989;311;76;0.244373;6.0614e-07\\
4.07;2.54824;0.0289544;442;46;0.104072;2.43038e-07\\
4.08;2.55824;0.0288104;955;39;0.0408377;8.78105e-08\\
}node [pos=0.8,anchor=south,font=\footnotesize,sloped,xshift = 1] {$L=3$};

\addplot [color=KITblue, line width=0.9pt, mark=none]table[x=delta, y=BER, col sep=semicolon,row sep=crcr] {
EbNo;EsNo;delta;decodedFrame;FrameErr;FER;BER\\
3.6;2.0782;0.036208;0;0;0;0.023831\\
3.65;2.1282;0.035391;0;0;0;0.021074\\
3.7;2.1782;0.034585;0;0;0;0.017897\\
3.75;2.2282;0.03379;0;0;0;0.014236\\
3.8;2.2782;0.033006;0;0;0;0.010133\\
3.85;2.3282;0.032232;0;0;0;0.0059247\\
3.9;2.3782;0.031469;0;0;0;0.0024221\\
3.95;2.4282;0.030716;0;0;0;0.00051977\\
4;2.4782;0.029975;0;0;0;3.7438e-05\\
4.05;2.5282;0.029244;0;0;0;5.1067e-07\\
4.1;2.5782;0.028524;0;0;0;7.1275e-10\\
};

\addplot [color=KITblue!40!KITred, dashed,line width=0.9pt, mark=o, mark options={solid,fill=white, mark size=1pt}]table[x=delta, y=BER, col sep=semicolon,row sep=crcr] {
EbNo;EsNo;delta;decodedFrame;FrameErr;FER;BER\\
3.5;1.97824;0.0378718;155;155;1;0.0281672\\
3.55;2.02824;0.0370344;156;156;1;0.0254679\\
3.6;2.07824;0.0362076;157;157;1;0.0218939\\
3.65;2.12824;0.0353912;157;157;1;0.0163763\\
3.675;2.15324;0.034987;156;156;1;0.0121655\\
3.7;2.17824;0.0345854;155;155;1;0.00753908\\
3.725;2.20324;0.0341865;157;157;1;0.00291679\\
3.75;2.22824;0.0337903;157;157;1;0.0003861\\
3.76;2.23824;0.0336325;158;150;0.949367;0.000143523\\
3.77;2.24824;0.0334752;162;118;0.728395;1.97862e-05\\
3.775;2.25324;0.0333967;178;97;0.544944;1.16495e-05\\
3.78;2.25824;0.0333183;200;67;0.335;3.14645e-06\\
3.785;2.26324;0.03324;290;50;0.172414;7.33677e-07\\
3.79;2.26824;0.0331618;593;40;0.0674536;2.45431e-07\\
3.795;2.27324;0.0330837;1517;31;0.0204351;7.35538e-08\\
}node [pos=0.8,anchor=south,font=\footnotesize,sloped,xshift = 1] {$L=5$};

\addplot [color=KITblue!40!KITred, line width=0.9pt, mark=none]table[x=delta, y=BER, col sep=semicolon,row sep=crcr] {
EbNo;EsNo;delta;decodedFrame;FrameErr;FER;BER\\
3.6;2.0782;0.036208;0;0;0;0.021686\\
3.61;2.0882;0.036043;0;0;0;0.020811\\
3.62;2.0982;0.03588;0;0;0;0.019858\\
3.63;2.1082;0.035716;0;0;0;0.018813\\
3.64;2.1182;0.035554;0;0;0;0.017661\\
3.65;2.1282;0.035391;0;0;0;0.016386\\
3.66;2.1382;0.035229;0;0;0;0.014969\\
3.67;2.1482;0.035068;0;0;0;0.013392\\
3.68;2.1582;0.034906;0;0;0;0.011645\\
3.69;2.1682;0.034746;0;0;0;0.0097307\\
3.7;2.1782;0.034585;0;0;0;0.0076802\\
3.71;2.1882;0.034426;0;0;0;0.0055768\\
3.72;2.1982;0.034266;0;0;0;0.0035757\\
3.73;2.2082;0.034107;0;0;0;0.0018993\\
3.74;2.2182;0.033948;0;0;0;0.00075789\\
3.75;2.2282;0.03379;0;0;0;0.0001966\\
3.76;2.2382;0.033633;0;0;0;2.7108e-05\\
3.77;2.2482;0.033475;0;0;0;1.5304e-06\\
3.78;2.2582;0.033318;0;0;0;2.572e-08\\
3.79;2.2682;0.033162;0;0;0;8.8016e-11\\
};

\addplot [color=KITred, dashed,line width=0.9pt, mark=o, mark options={solid,fill=white, mark size=1pt}]table[x=delta, y=BER, col sep=semicolon,row sep=crcr] {
EbNo;EsNo;delta;decodedFrame;FrameErr;FER;BER\\
3.5;1.97824;0.0378718;157;157;1;0.0280585\\
3.55;2.02824;0.0370344;155;155;1;0.0250707\\
3.6;2.07824;0.0362076;156;156;1;0.0198756\\
3.65;2.12824;0.0353912;157;157;1;0.0068483\\
3.675;2.15324;0.034987;173;132;0.763006;0.000435564\\
3.685;2.16324;0.0348261;215;75;0.348837;4.66922e-05\\
3.69;2.16824;0.0347458;302;55;0.182119;2.12069e-05\\
3.695;2.17324;0.0346655;607;40;0.0658979;3.8957e-06\\
3.7;2.17824;0.0345854;1809;30;0.0165837;3.54379e-07\\
3.705;2.18324;0.0345055;8704;31;0.00356158;8.73479e-08\\
}node [pos=0.8,anchor=south,font=\footnotesize,sloped,xshift = 1] {$L=7$};

\addplot [color=KITred, line width=0.9pt, mark=none]table[x=delta, y=BER, col sep=semicolon,row sep=crcr] {
EbNo;EsNo;delta;decodedFrame;FrameErr;FER;BER\\
3.6;2.0782;0.036208;0;0;0;0.020031\\
3.605;2.0832;0.036125;0;0;0;0.019314\\
3.61;2.0882;0.036043;0;0;0;0.018521\\
3.615;2.0932;0.035962;0;0;0;0.017637\\
3.62;2.0982;0.03588;0;0;0;0.016645\\
3.625;2.1032;0.035798;0;0;0;0.015524\\
3.63;2.1082;0.035716;0;0;0;0.014251\\
3.635;2.1132;0.035635;0;0;0;0.012799\\
3.64;2.1182;0.035554;0;0;0;0.011145\\
3.645;2.1232;0.035472;0;0;0;0.0092789\\
3.65;2.1282;0.035391;0;0;0;0.0072207\\
3.655;2.1332;0.03531;0;0;0;0.0050595\\
3.66;2.1382;0.035229;0;0;0;0.002996\\
3.665;2.1432;0.035148;0;0;0;0.0013443\\
3.67;2.1482;0.035068;0;0;0;0.00037957\\
3.675;2.1532;0.034987;0;0;0;4.9774e-05\\
3.68;2.1582;0.034906;0;0;0;1.9107e-06\\
3.685;2.1632;0.034826;0;0;0;1.1211e-08\\
3.69;2.1682;0.034746;0;0;0;4.1711e-12\\
};

\addplot [color=KITred!70!KITorange, dashed,line width=0.9pt, mark=o, mark options={solid,fill=white, mark size=1pt}]table[x=delta, y=BER, col sep=semicolon,row sep=crcr] {
EbNo;EsNo;delta;decodedFrame;FrameErr;FER;BER\\
3.1;1.57824;0.0449425;218;218;1;0.0413092\\
3.2;1.67824;0.0431136;221;221;1;0.0385689\\
3.3;1.77824;0.0413253;221;221;1;0.0356262\\
3.4;1.87824;0.0395778;217;217;1;0.032249\\
3.5;1.97824;0.0378718;214;214;1;0.0279655\\
3.55;2.02824;0.0370344;217;217;1;0.0248136\\
3.6;2.07824;0.0362076;220;220;1;0.015154\\
3.625;2.10324;0.0357981;195;107;0.548718;0.00149521\\
3.64;2.11824;0.0355536;577;33;0.0571924;5.78427e-05\\
3.65;2.12824;0.0353912;8681;30;0.00345582;1.67973e-06\\
3.655;2.13324;0.0353102;72971;30;0.000411122;1.10239e-07\\
};

 \addplot [color=KITred!70!KITorange, line width=0.9pt, mark = none]
table[x=delta, y=BER, col sep=semicolon,row sep=crcr]{%
EbNo;EsNo;delta;decodedFrame;FrameErr;FER;BER\\
3.6;2.0782;0.036208;0;0;0;0.016867\\
3.601;2.0792;0.036191;0;0;0;0.016497\\
3.602;2.0802;0.036175;0;0;0;0.016103\\
3.603;2.0812;0.036158;0;0;0;0.015685\\
3.604;2.0822;0.036142;0;0;0;0.015238\\
3.605;2.0832;0.036125;0;0;0;0.014762\\
3.606;2.0842;0.036109;0;0;0;0.014252\\
3.607;2.0852;0.036093;0;0;0;0.013705\\
3.608;2.0862;0.036076;0;0;0;0.01312\\
3.609;2.0872;0.03606;0;0;0;0.012492\\
3.61;2.0882;0.036043;0;0;0;0.011817\\
3.611;2.0892;0.036027;0;0;0;0.011094\\
3.612;2.0902;0.036011;0;0;0;0.01032\\
3.613;2.0912;0.035994;0;0;0;0.0094928\\
3.614;2.0922;0.035978;0;0;0;0.0086129\\
3.615;2.0932;0.035962;0;0;0;0.0076833\\
3.616;2.0942;0.035945;0;0;0;0.0067104\\
3.617;2.0952;0.035929;0;0;0;0.0057066\\
3.618;2.0962;0.035912;0;0;0;0.0046909\\
3.619;2.0972;0.035896;0;0;0;0.0036915\\
3.62;2.0982;0.03588;0;0;0;0.0027454\\
3.621;2.0992;0.035863;0;0;0;0.0018963\\
3.622;2.1002;0.035847;0;0;0;0.0011885\\
3.623;2.1012;0.035831;0;0;0;0.00065522\\
3.624;2.1022;0.035814;0;0;0;0.0003049\\
3.625;2.1032;0.035798;0;0;0;0.00011349\\
3.626;2.1042;0.035782;0;0;0;3.1536e-05\\
3.627;2.1052;0.035765;0;0;0;6.0005e-06\\
3.628;2.1062;0.035749;0;0;0;7.0265e-07\\
3.629;2.1072;0.035733;0;0;0;4.4459e-08\\
3.63;2.1082;0.035716;0;0;0;1.2977e-09\\
};

 \addplot [color=KITorange,  line width=0.9pt, mark = none]
table[x=delta, y=BER, col sep=semicolon,row sep=crcr]{%
EbNo;EsNo;delta;decodedFrame;FrameErr;FER;BER\\
3.5219;2.0001;0.037504;0;0;0;0.026533\\
3.5319;2.0101;0.037336;0;0;0;0.025901\\
3.5419;2.0201;0.037169;0;0;0;0.025195\\
3.5519;2.0301;0.037003;0;0;0;0.024371\\
3.5619;2.0401;0.036837;0;0;0;0.023315\\
3.5712;2.0494;0.036683;0;0;0;0.021238\\
3.5713;2.0496;0.03668;0;0;0;0.0000000000001\\
}node [pos=0.38,anchor=south,font=\footnotesize,sloped,yshift = -13] {$L=1000$};

\node[ellipse,draw,KITred!70!KITorange, thick, dashed, anchor = center,at={(axis cs:0.0356,2e-5)},minimum width = 0.2cm, 
    minimum height = 0.1cm] (e2)  {};
\node[font=\footnotesize, rotate = 90, at={(axis cs:0.038,1e-4)}] (e1) {\textcolor{KITred!70!KITorange}{$L=10$}};
\draw[dashed, thick,KITred!70!KITorange] (e2)--(e1);
\end{axis}
\end{tikzpicture} 

%% file: img/nu_7_t_1_e_1.tex
\begin{tikzpicture}
\pgfplotsset{grid style={dashed, gray}}
\pgfplotsset{every tick label/.append style={font=\footnotesize}}
\begin{axis}[%
width=.9\columnwidth,
height=5.5cm,
xmin=0.001,
xmax=0.025,
ymode=log,
ymin=1e-6,
ymax=0.1,
yminorticks,
axis background/.style={fill=white, mark size=1pt},
xmajorgrids,
ymajorgrids,
yminorgrids,
ytick={0.1,0.01,0.001,1e-4,1e-5,1e-6,1e-7,1e-8},
ylabel={BER},
xlabel={Cross-over probability $p$},
label style={font=\small},
legend cell align={left},
legend style={anchor =north west, at={(0,1)},draw=black, fill opacity=1, text opacity = 1,legend columns=1, row sep = 0pt,font=\small}
]

\addplot [color=black, dashed, line width=0.9pt, mark=none]table[x=delta, y=BER, col sep=semicolon,row sep=crcr]{
EbNo;EsNo;delta;decodedFrame;FrameErr;FER;MC_rate;BER\\
4.2;0;0;0;0;0;0;1e-100\\
};
\addlegendentry{DE}

\addplot [color=black,  line width=0.9pt, mark=o, mark options={solid,fill=white, mark size=1pt}]table[x=delta, y=BER, col sep=semicolon,row sep=crcr] {
EbNo;EsNo;delta;decodedFrame;FrameErr;FER;MC_rate;BER\\
4.2;0;0;0;0;0;0;1e-100\\
};
\addlegendentry{MF iBDD}

\addplot [color=KITblue, line width=0.9pt, mark=none]table[x=delta, y=BER, col sep=semicolon,row sep=crcr] {
EbNo;EsNo;delta;decodedFrame;FrameErr;FER;MC_rate;BER\\
4.2;0;0;0;0;0;0;1e-100\\
};
\addlegendentry{Staircase}

\addplot [color=Apricot, line width=0.9pt, mark=none]table[x=delta, y=BER, col sep=semicolon,row sep=crcr] {
EbNo;EsNo;delta;decodedFrame;FrameErr;FER;MC_rate;BER\\
4.2;0;0;0;0;0;0;1e-100\\
};
\addlegendentry{Half ESC}

\addplot [color=KITred, line width=0.9pt, mark=none]table[x=delta, y=BER, col sep=semicolon,row sep=crcr] {
EbNo;EsNo;delta;decodedFrame;FrameErr;FER;MC_rate;BER\\
4.2;0;0;0;0;0;0;1e-100\\
};
\addlegendentry{ESC}

\addplot [color=KITblue, dashed, line width=0.9pt, mark=none]table[x=delta, y=BER, col sep=semicolon,row sep=crcr]{
EbNo;EsNo;delta;decodedFrame;FrameErr;FER;BER\\
4;2.8712;0.024522;0;0;0;0.011967\\
4.05;2.9212;0.023876;0;0;0;0.010977\\
4.1;2.9712;0.02324;0;0;0;0.010008\\
4.15;3.0212;0.022615;0;0;0;0.0090609\\
4.2;3.0712;0.022002;0;0;0;0.0081396\\
4.25;3.1212;0.021398;0;0;0;0.0072467\\
4.3;3.1712;0.020806;0;0;0;0.0063857\\
4.35;3.2212;0.020224;0;0;0;0.0055602\\
4.4;3.2712;0.019653;0;0;0;0.0047746\\
4.45;3.3212;0.019092;0;0;0;0.0040333\\
4.5;3.3712;0.018542;0;0;0;0.0033416\\
4.55;3.4212;0.018002;0;0;0;0.0027049\\
4.6;3.4712;0.017473;0;0;0;0.0021292\\
4.65;3.5212;0.016954;0;0;0;0.0016204\\
4.7;3.5712;0.016446;0;0;0;0.0011841\\
4.75;3.6212;0.015948;0;0;0;0.0008244\\
4.8;3.6712;0.01546;0;0;0;0.00054268\\
4.85;3.7212;0.014982;0;0;0;0.00033583\\
5;3.8712;0.013609;0;0;0;5.6114e-05\\
5.15;4.0212;0.012325;0;0;0;7.0786e-06\\
5.3;4.1712;0.011128;0;0;0;8.7877e-07\\
5.6;4.4712;0.0089831;0;0;0;1.5332e-08\\
};

\addplot [color=Apricot, dashed, line width=0.9pt, mark=none]table[x=delta, y=BER, col sep=semicolon,row sep=crcr]{
EbNo;EsNo;delta;decodedFrame;FrameErr;FER;BER\\
4;2.873;0.024498;0;0;0;0.011226\\
4.1;2.973;0.023217;0;0;0;0.0080031\\
4.2;3.073;0.02198;0;0;0;0.004489\\
4.3;3.173;0.020785;0;0;0;0.0014089\\
4.4;3.273;0.019632;0;0;0;0.00026397\\
4.5;3.373;0.018522;0;0;0;5.1849e-05\\
4.6;3.473;0.017454;0;0;0;1.1258e-05\\
4.7;3.573;0.016428;0;0;0;2.6071e-06\\
4.8;3.673;0.015442;0;0;0;6.3108e-07\\
};

\addplot [color=KITred, dashed, line width=0.9pt, mark=none]table[x=delta, y=BER, col sep=semicolon,row sep=crcr]{
EbNo;EsNo;delta;decodedFrame;FrameErr;FER;BER\\
4;2.8712;0.024522;0;0;0;0.01362\\
4.01;2.8812;0.024392;0;0;0;0.013183\\
4.02;2.8912;0.024262;0;0;0;0.012712\\
4.03;2.9012;0.024133;0;0;0;0.012198\\
4.04;2.9112;0.024004;0;0;0;0.011625\\
4.05;2.9212;0.023876;0;0;0;0.010975\\
4.06;2.9312;0.023748;0;0;0;0.010218\\
4.07;2.9412;0.02362;0;0;0;0.0093069\\
4.08;2.9512;0.023493;0;0;0;0.00816\\
4.09;2.9612;0.023366;0;0;0;0.006612\\
4.1;2.9712;0.02324;0;0;0;0.0042536\\
4.11;2.9812;0.023114;0;0;0;0.000622\\
4.12;2.9912;0.022989;0;0;0;5.717e-05\\
4.13;3.0012;0.022864;0;0;0;2.3037e-05\\
4.14;3.0112;0.02274;0;0;0;1.1871e-05\\
4.15;3.0212;0.022615;0;0;0;6.7603e-06\\
4.16;3.0312;0.022492;0;0;0;4.0677e-06\\
4.17;3.0412;0.022369;0;0;0;2.5346e-06\\
4.18;3.0512;0.022246;0;0;0;1.6177e-06\\
4.19;3.0612;0.022124;0;0;0;1.0507e-06\\
4.2;3.0712;0.022002;0;0;0;6.9153e-07\\
4.21;3.0812;0.02188;0;0;0;4.5987e-07\\
4.22;3.0912;0.021759;0;0;0;3.0837e-07\\
4.23;3.1012;0.021638;0;0;0;2.0818e-07\\
4.24;3.1112;0.021518;0;0;0;1.4133e-07\\
};

\addplot [color=KITblue,  line width=0.9pt, mark=o, mark options={solid,KITblue,fill=white, mark size=1pt}]table[x=delta, y=BER, col sep=semicolon,row sep=crcr] {
EbNo;EsNo;delta;decodedFrame;FrameErr;FER;MC_rate;BER\\
3.5;2.37296;0.0315489;427;427;1;0;0.0224777\\
3.9;2.77296;0.025822;426;426;1;0;0.013817\\
4.3;3.17296;0.0207847;426;426;1;0;0.00610124\\
4.5;3.37296;0.0185223;427;427;1;0;0.00322532\\
4.7;3.57296;0.0164277;427;427;1;0;0.00132815\\
4.9;3.77296;0.0144977;426;426;1;0;0.00053054\\
5.1;3.97296;0.0127281;450;414;0.92;0;0.000210681\\
5.3;4.17296;0.0111138;543;388;0.714549;0;8.61842e-05\\
5.5;4.37296;0.00964924;657;368;0.560122;0;5.00002e-05\\
5.7;4.57296;0.00832803;1046;345;0.329828;0;2.5337e-05\\
5.9;4.77296;0.00714317;1772;324;0.182844;0;1.20254e-05\\
6.1;4.97296;0.00608719;3652;312;0.0854326;0;5.32229e-06\\
6.22;5.09296;0.00551218;4491;309;0.0688043;0;4.10548e-06\\
6.34;5.21296;0.00497896;8737;305;0.034909;0;2.04909e-06\\
6.46;5.33296;0.00448571;11786;301;0.0255388;0;1.47601e-06\\
6.58;5.45296;0.00403063;18746;300;0.0160034;0;9.11478e-07\\
};

\addplot [color=Apricot,  line width=0.9pt, mark=o, mark options={solid,fill=white, mark size=1pt}]table[x=delta, y=BER, col sep=semicolon,row sep=crcr] {
EbNo;EsNo;delta;decodedFrame;FrameErr;FER;MC_rate;BER\\
3.5;2.37296;0.0315489;225;225;1;0;0.0247274\\
3.7;2.57296;0.0285992;227;227;1;0;0.0197712\\
3.9;2.77296;0.025822;227;227;1;0;0.0145451\\
4.1;2.97296;0.0232174;227;227;1;0;0.00817517\\
4.2;3.07296;0.0219796;227;227;1;0;0.00503995\\
4.3;3.17296;0.0207847;226;226;1;0;0.00254949\\
4.4;3.27296;0.0196323;226;226;1;0;0.0011401\\
4.5;3.37296;0.0185223;233;226;0.969957;0;0.000535076\\
4.6;3.47296;0.0174542;233;212;0.909871;0;0.000262947\\
4.7;3.57296;0.0164277;255;198;0.776471;0;0.000130204\\
4.8;3.67296;0.0154424;278;173;0.622302;0;6.61708e-05\\
4.9;3.77296;0.0144977;355;165;0.464789;0;4.30564e-05\\
5.1;3.97296;0.0127281;495;125;0.252525;0;1.82543e-05\\
5.3;4.17296;0.0111138;701;118;0.168331;0;1.12637e-05\\
5.52;4.39296;0.00951078;1294;109;0.0842349;0;4.99246e-06\\
5.7;4.57296;0.00832803;2205;105;0.047619;0;2.84683e-06\\
5.88;4.75296;0.00725573;3466;103;0.0297173;0;1.72176e-06\\
6.06;4.93296;0.00628845;6265;100;0.0159617;0;8.87385e-07\\
};

\addplot [color=KITred, line width=0.9pt, mark=o, mark options={solid,fill=white, mark size=1pt}]table[x=delta, y=BER, col sep=semicolon,row sep=crcr] {
EbNo;EsNo;delta;decodedFrame;FrameErr;FER;MC_rate;BER\\
3.5;2.37296;0.0315489;427;427;1;0;0.0270996\\
3.7;2.57296;0.0285992;427;427;1;0;0.0225124\\
3.9;2.77296;0.025822;427;427;1;0;0.0172098\\
4.1;2.97296;0.0232174;427;427;1;0;0.00882787\\
4.2;3.07296;0.0219796;438;422;0.96347;0;0.00345621\\
4.3;3.17296;0.0207847;481;408;0.848233;0;0.000866363\\
4.4;3.27296;0.0196323;743;366;0.492598;0;0.000154726\\
4.5;3.37296;0.0185223;1361;331;0.243204;0;3.15515e-05\\
4.6;3.47296;0.0174542;3609;315;0.0872818;0;5.87709e-06\\
4.66;3.53296;0.0168334;6203;306;0.049331;0;2.76591e-06\\
4.72;3.59296;0.0162274;10466;305;0.029142;0;1.72268e-06\\
4.78;3.65296;0.0156362;20255;302;0.0149099;0;7.34478e-07\\
};

\end{axis}
\end{tikzpicture} 

%% file: img/nu_9_t_2_e_1.tex
\begin{tikzpicture}
\pgfplotsset{grid style={dashed, gray}}
\pgfplotsset{every tick label/.append style={font=\footnotesize}}
\begin{axis}[%
width=.9\columnwidth,
height=5.5cm,
xmin=0.014,
xmax=0.02,
ymode=log,
ymin=1e-7,
ymax=0.1,
yminorticks,
axis background/.style={fill=white, mark size=1pt},
xmajorgrids,
ymajorgrids,
yminorgrids,
label style={font=\small},
xtick={0.012,0.014,0.016,0.018,0.02},
ytick={0.1,0.01,0.001,1e-4,1e-5,1e-6,1e-7,1e-8},
ylabel={BER},
xlabel={Cross-over probability $p$},
legend cell align={left},
legend style={anchor =north west, at={(0,1)},draw=black, fill opacity=1, text opacity = 1,legend columns=1, row sep = 0pt,font=\small}
]

\addplot [color=black, dashed, line width=0.9pt, mark=none]table[x=delta, y=BER, col sep=semicolon,row sep=crcr]{
EbNo;EsNo;delta;decodedFrame;FrameErr;FER;MC_rate;BER\\
4.2;0;0;0;0;0;0;1e-100\\
};
\addlegendentry{DE}

\addplot [color=black,  line width=0.9pt, mark=o, mark options={solid,fill=white, mark size=1pt}]table[x=delta, y=BER, col sep=semicolon,row sep=crcr] {
EbNo;EsNo;delta;decodedFrame;FrameErr;FER;MC_rate;BER\\
4.2;0;0;0;0;0;0;1e-100\\
};
\addlegendentry{MF iBDD}

\addplot [color=KITblue, line width=0.9pt, mark=none]table[x=delta, y=BER, col sep=semicolon,row sep=crcr] {
EbNo;EsNo;delta;decodedFrame;FrameErr;FER;MC_rate;BER\\
4.2;0;0;0;0;0;0;1e-100\\
};
\addlegendentry{Staircase}

\addplot [color=Apricot, line width=0.9pt, mark=none]table[x=delta, y=BER, col sep=semicolon,row sep=crcr] {
EbNo;EsNo;delta;decodedFrame;FrameErr;FER;MC_rate;BER\\
4.2;0;0;0;0;0;0;1e-100\\
};
\addlegendentry{Half ESC}

\addplot [color=KITred, line width=0.9pt, mark=none]table[x=delta, y=BER, col sep=semicolon,row sep=crcr] {
EbNo;EsNo;delta;decodedFrame;FrameErr;FER;MC_rate;BER\\
4.2;0;0;0;0;0;0;1e-100\\
};
\addlegendentry{ESC}
\addplot [color=KITblue,  line width=0.9pt, mark=o, mark options={solid,KITblue,fill=white, mark size=1pt}]table[x=delta, y=BER, col sep=semicolon,row sep=crcr] {
EbNo;EsNo;delta;decodedFrame;FrameErr;FER;MC_rate;BER\\
4.2;3.2309;0.0201118;218;218;1;0;0.0135025\\
4.23;3.2609;0.019769;163;163;1;0;0.0127424\\
4.26;3.2909;0.01943;163;163;1;0;0.011985\\
4.29;3.3209;0.0190949;163;163;1;0;0.0110622\\
4.32;3.3509;0.0187635;163;163;1;0;0.0100234\\
4.35;3.3809;0.0184359;163;163;1;0;0.00885003\\
4.38;3.4109;0.0181121;163;163;1;0;0.00698717\\
4.4;3.4309;0.0178983;221;221;1;0;0.00551328\\
4.45;3.4809;0.0173712;224;213;0.950893;0;0.00202803\\
4.5;3.5309;0.0168544;262;149;0.568702;0;0.000356224\\
4.55;3.5809;0.016348;376;57;0.151596;0;3.2172e-05\\
4.6;3.6309;0.0158518;993;35;0.0352467;0;4.40181e-06\\
4.63;3.6609;0.015559;2666;36;0.0135034;0;9.43019e-07\\
4.66;3.6909;0.0152699;10477;30;0.00286342;0;1.72678e-07\\
4.7;3.7309;0.0148901;39033;30;0.00076858;0;1.91368e-08\\
};

\addplot [color=KITblue, dashed, line width=0.9pt, mark=none]table[x=delta, y=BER, col sep=semicolon,row sep=crcr]{
EbNo;EsNo;delta;decodedFrame;FrameErr;FER;BER\\
4.2;3.2306;0.020115;0;0;0;0.012628\\
4.22;3.2506;0.019887;0;0;0;0.012102\\
4.24;3.2706;0.019659;0;0;0;0.011562\\
4.26;3.2906;0.019434;0;0;0;0.011007\\
4.28;3.3106;0.01921;0;0;0;0.010435\\
4.3;3.3306;0.018988;0;0;0;0.0098421\\
4.32;3.3506;0.018767;0;0;0;0.0092248\\
4.34;3.3706;0.018548;0;0;0;0.0085768\\
4.36;3.3906;0.018331;0;0;0;0.0078887\\
4.38;3.4106;0.018116;0;0;0;0.0071424\\
4.4;3.4306;0.017902;0;0;0;0.0062877\\
4.42;3.4506;0.01769;0;0;0;4.1558e-07\\
4.44;3.4706;0.017479;0;0;0;8.7794e-12\\
};

\addplot [color=Apricot, line width=0.9pt, mark=o, mark options={solid,Apricot,fill=white, mark size=1pt}]table[x=delta, y=BER, col sep=semicolon,row sep=crcr] {
EbNo;EsNo;delta;decodedFrame;FrameErr;FER;MC_rate;BER\\
4.2;3.2309;0.0201118;220;220;1;0;0.0164732\\
4.23;3.2609;0.019769;163;163;1;0;0.0155644\\
4.26;3.2909;0.01943;163;163;1;0;0.0144442\\
4.29;3.3209;0.0190949;163;163;1;0;0.0127116\\
4.32;3.3509;0.0187635;168;161;0.958333;0;0.00942199\\
4.38;3.4109;0.0181121;232;136;0.586207;0;0.00292196\\
4.4;3.4309;0.0178983;286;91;0.318182;0;0.00114096\\
4.4;3.4309;0.0178983;261;106;0.40613;0;0.00122876\\
4.45;3.4809;0.0173712;570;40;0.0701754;0;6.87175e-05\\
4.5;3.5309;0.0168544;3756;30;0.00798722;0;2.42768e-06\\
4.53;3.5609;0.0165493;17296;31;0.00179232;0;2.94911e-07\\
4.56;3.5909;0.0162479;82317;30;0.000364445;0;1.75916e-08\\
};

\addplot [color=Apricot, dashed, line width=0.9pt, mark=none]table[x=delta, y=BER, col sep=semicolon,row sep=crcr]{
EbNo;EsNo;delta;decodedFrame;FrameErr;FER;BER\\
4.2;3.2306;0.020115;0;0;0;0.016437\\
4.22;3.2506;0.019887;0;0;0;0.015989\\
4.24;3.2706;0.019659;0;0;0;0.015518\\
4.26;3.2906;0.019434;0;0;0;0.015011\\
4.28;3.3106;0.01921;0;0;0;0.01441\\
4.3;3.3306;0.018988;0;0;0;0.011461\\
4.32;3.3506;0.018767;0;0;0;1.1011e-09\\
4.34;3.3706;0.018548;0;0;0;1.1232e-11\\
4.36;3.3906;0.018331;0;0;0;1.13e-13\\
};

\addplot [color=KITred, line width=0.9pt, mark=o, mark options={solid,fill=white, mark size=1pt}]table[x=delta, y=BER, col sep=semicolon,row sep=crcr] {
EbNo;EsNo;delta;decodedFrame;FrameErr;FER;MC_rate;BER\\
4.2;3.2309;0.0201118;221;221;1;0;0.0181523\\
4.23;3.2609;0.019769;163;163;1;0;0.0172261\\
4.26;3.2909;0.01943;164;163;0.993902;0;0.0153019\\
4.3;3.3309;0.018984;224;209;0.933036;0;0.01067\\
4.32;3.3509;0.0187635;185;149;0.805405;0;0.00756041\\
4.35;3.3809;0.0184359;271;131;0.483395;0;0.00371684\\
4.38;3.4109;0.0181121;569;115;0.202109;0;0.00110779\\
4.4;3.4309;0.0178983;546;47;0.0860806;0;0.000549884\\
4.45;3.4809;0.0173712;2867;31;0.0108127;0;3.19218e-05\\
4.5;3.5309;0.0168544;30882;30;0.00097144;0;1.83703e-06\\
4.53;3.5609;0.0165493;243736;30;0.000123084;0;6.76341e-08\\
};

\addplot [color=KITred, dashed, line width=0.9pt, mark=none]table[x=delta, y=BER, col sep=semicolon,row sep=crcr]{
EbNo;EsNo;delta;decodedFrame;FrameErr;FER;BER\\
4.2;3.2306;0.020115;0;0;0;0.018533\\
4.22;3.2506;0.019887;0;0;0;0.018202\\
4.24;3.2706;0.019659;0;0;0;0.017849\\
4.26;3.2906;0.019434;0;0;0;0.017322\\
4.28;3.3106;0.01921;0;0;0;5.1025e-12\\
};

\end{axis}
\end{tikzpicture} 